\documentclass[12pt,a4paper]{article}
\usepackage{amsmath,amssymb,mathrsfs,mathtools,framed,esint}
\usepackage[colorlinks]{hyperref}
\usepackage{color,soul}
\usepackage[all]{xy}

\newcommand{\rem}[1]{}
\newcommand{\bc}{{\boldsymbol{c}}}

\newcommand{\de}{{\rm d}}

\newcommand{\z}{{\mathbf{z}}}

\newcommand{\bq}{{\mathbf{q}}}
\newcommand{\ba}{{\boldsymbol{a}}}

\newcommand{\bM}{{\mathbf{M}}}

\newcommand{\bx}{{\mathbf{x}}}
\newcommand{\bv}{{\mathbf{v}}}
\newcommand{\bw}{{\boldsymbol{w}}}
\newcommand{\bb}{{\boldsymbol{b}}}
\newcommand{\bA}{{\mathbf{A}}}

\newcommand{\bK}{{\mathbf{K}}}
\newcommand{\bV}{{\boldsymbol{V}}}
\newcommand{\bE}{{\mathbf{E}}}
\newcommand{\bB}{{\mathbf{B}}}
\newcommand{\bJ}{{\mathbf{J}}}

\newcommand{\bu}{{\boldsymbol{u}}}
\newcommand{\bU}{ {\boldsymbol{U}} }

\newcommand{\bcalX}{ {\boldsymbol{\mathcal{X}}} }

\newcommand{\bY}{{\boldsymbol{Y}}}

\newcommand{\bz}{{\mathbf{z}}}

\newcommand{\bcalXi}{{\boldsymbol{\Xi}}}
\newcommand{\bxi}{{\boldsymbol{\xi}}}

\newcommand{\vp}{{v_\parallel}}
\newcommand{\pa}{\partial}
\newcommand{\et}{{\boldsymbol\eta}}
\newcommand{\ps}{{\boldsymbol\psi}}

\newcommand{\beq}{\begin{equation}}
\newcommand{\eeq}{\end{equation}}
\newcommand{\ben}{\begin{eqnarray}}
\newcommand{\een}{\end{eqnarray}}

\begin{document}

\title{A low-frequency variational model for energetic particle effects in the pressure-coupling scheme}


\author{Alexander R.~D. Close$^1$, Joshua W. Burby$^2$, Cesare Tronci$^1$\\
\it\footnotesize $^1$Department of Mathematics, University of Surrey, Guildford GU2 7XH, United Kingdom\\
\it\footnotesize $^2$Courant Institute of Mathematical Sciences, New York University, New York 10012, United States
}

\date{}

\maketitle

\begin{abstract}  
Energetic particle effects in magnetic confinement fusion devices are commonly studied by hybrid kinetic-fluid simulation codes whose underlying continuum evolution equations often lack the correct energy balance. While two different kinetic-fluid coupling options are available (current-coupling and pressure-coupling), this paper applies the Euler-Poincar\'e variational approach to formulate a new conservative hybrid model in the pressure-coupling scheme. In our case the kinetics of the energetic particles are described by guiding center theory. The interplay between the Lagrangian fluid paths with phase space particle trajectories reflects an intricate variational structure which can be approached by letting the 4-dimensional guiding center trajectories evolve in the full 6-dimensional phase space. Then, the redundant perpendicular velocity is  integrated out to recover a four-dimensional description. A second equivalent variational approach is also reported, which involves the use of phase space Lagrangians. Not only do these variational structures confer on the new  model a correct energy balance, but also they produce a cross-helicity invariant which is lost in the other pressure-coupling schemes  reported in the literature.
\end{abstract}

\bigskip

{\footnotesize
\tableofcontents
\newpage}

\section{Nonlinear hybrid models for energetic particles}

The interaction of bulk fluid plasmas with energetic particles populating magnetic confinement devices has been widely studied over several decades. {After early work} on the linear regime \cite{Chen,Coppi}, nonlinear hybrid kinetic-fluid models started appearing in the early 90's \cite{Cheng,ParkEtAl} along with two coupling options: {a} current-coupling scheme (CCS) and a pressure-coupling scheme (PCS). Although these two variants involve different levels of approximations, they were considered essentially equivalent \cite{ParkEtAl} and both of them have been widely implemented in numerical simulations over the years \cite{FuParketAl,Kim,Hou,Todo,Pei}. {Our purview in this article is} the nonlinear models underlying these two coupling options, {but} giving special emphasis to the PCS.

Recently, a thorough investigation of the continuum equations underlying these numerical schemes has led to relevant considerations concerning the energy balance \cite{TrTaCaMo,BuTr} in the ideal (dissipationless) limit. {It was shown in particular} that the correct energy balance can be completely lost, {with this loss depending} essentially on two points: 1) the particular coupling option that is adopted (pressure-coupling or current-coupling) and 2) the description {used to} model energetic particle kinetics (full-orbit trajectories, drift-kinetic or gyrokinetic approximation). 

First, we consider current-coupling schemes. {It was shown} \cite{BuTr} {in this context that} the correct energy balance is dramatically lost in the ideal continuum equations underlying the MEGA code \cite{Pei,Todo,ToSa,ToSaWaWaHo}, whose energetic particle kinetics {are} described by the guiding center approximation. Similar considerations also apply when considering gyrokinetic theory \cite{BeDeCh,BuTr}. {When full-orbit Vlasov kinetics {are} employed, the current-coupling scheme currently adopted by the HYM code} \cite{BeGoFrTrCr,BelovaPark} {has so far emerged} as the only model for energetic particles (to our knowledge) that possesses the correct energy balance \cite{Tronci2010,HoTr2011}. {It is this feature that has} recently enabled {the proof of} the existence of global-in{-}time weak solutions for the resistive equations {of} the HYM code \cite{ChSuTr}.

The situation is more involved for pressure-coupling schemes{; the} extra approximations that are performed on the (fully nonlinear) continuum equations break the exact energy balance no matter the description adopted for the energetic particle kinetics.
This scenario shows not only that the CCS and the PCS are far from being equivalent{---}thereby disproving a common{ly held} belief \cite{PaBeFuTaStSu,ParkEtAl}{---}but also that the physical consistency {of available} codes may be undermined in the fully nonlinear regime. Indeed, {while} considering the PCS with Vlasov kinetics \cite{Kim1}, reference \cite{TrTaCaMo} showed that spurious high-frequency instabilities are triggered by unphysical energy sources{,} thereby affecting the physical reliability of the continuum model underlying the simulation code. 
A {potential} criticism {of} this point is that high-frequency instabilities are irrelevant in fusion devices where the processes of interest take place at frequencies well below the cyclotron frequency of the energetic particles. However, this argument {relates} strictly to spectral stability and tends to ignore the possible nonlinear behavior in which unstable high-frequency modes can couple to (and destabilize) low-frequen{cy} waves. {Another criticism might} be that this last situation can occur in highly turbulent (strongly nonlinear) phenomena that rarely take place in magnetic confinement fusion. Even if this point is accepted, then the whole reason to implement simulation codes based on fully nonlinear dynamics would fail to hold{,} thereby isolating linear models \cite{Chen,Bri94} as much more reliable than their nonlinear variants currently used in several studies.

{If the necessity of implementing fully nonlinear models is insisted upon, it may be tempting} to keep the available nonlinear models (lacking energy balance) and filter instabilities by adding several types of extra dissipation in{to} the model (including diffusion and hyper-diffusion) in the hope {of modeling} collisional effects that could perhaps correspond to real physical phenomena. {An alternative} direction is to identify the physical force terms that need to be added to the continuum equations in order to restore energy balance. This second strategy is precisely the program recently followed in the work stream \cite{BuTr,HoTr2011,MoTaTr2014-1,MoTaTr2014,Tronci2010,TrTaCaMo} to which this paper belongs. In this context, standard viscous and resistive terms can be added {\em a posteriori}, after the ideal (energy-conserving) model has been formulated.

In \cite{HoTr2011,Tronci2010}, the energy balance was restored in the ideal (dissipationless) equations by exploiting Hamiltonian and variational methods that have been used over the decades in several fields. In plasma theory {these} led to Littlejohn's discovery of guiding center theory \cite{Littlejohn} and its gyrokinetic extension \cite{BrHa}. {The} techniques have branched out in{to} several methods: noncanonical Hamiltonian structures \cite{Morrison2005,MorrisonGreene}; Lie series expansions \cite{BrHa}; variational principles for Eulerian dynamics (Euler-Poincar\'e) \cite{CeMaHo}. The same methods are {now} being exploited for the formulation of energy conserving particle-in-cell codes \cite{Krauss}.

We adopt the {last of these techniques in this paper in order to} formulate a new energy-conserving continuum model for the PCS, describing energetic particles by the guiding center theory ({of which a non-conservative version is} currently used in the NIMROD code \cite{Hou,Kim,KimEtAl,TaBrKi,TaBrKi2}). The plan is as follows. Upon considering full Vlasov kinetics, Section \ref{sec:Vlasov-PCS} reviews the formulation of the PCS model in both the non-conservative and the conservative variants. In Section \ref{non-cons}, the non-conservative PCS model from the literature is derived by making approximations on the CCS, which is known to be energy-conserving in this case. Then, Section \ref{sec:var1} reviews the Euler-Poincar\'e variational approach to the conservative PCS model and Section \ref{sec:EP} generalizes this construction to arbitrary action principles possessing the same Euler-Poincar\'e structure.

{T}he case of guiding center motion is considered in {S}ection \ref{sec:2}. Based on a recent conservative variant of the CCS model \cite{BuTr}, {S}ection \ref{sec:NC-PCS} illustrates the strategy followed in the literature to derive non-conservative PCS models. The new{ly} conservative PCS system is then {first} formulated in {S}ection \ref{sec:newmod}, where the novel idea is to embed the guiding center phase space within the full 6-dimensional phase space in such a way that the Lagrangian fluid paths can be naturally coupled to the particle paths in phase space. Thus, this section succeeds in deriving the new model using an extension of the variational method applied in Ref.\,\cite{HoTr2011} to derive a full-orbit PCS model. While {this procedure} introduces a redundancy, {it is eventually} projected out by integrating {over} the perpendicular velocity coordinates. The details of this treatment are relegated to the Appendix. 
{An alternative derivation of the same newly conservative PCS model is then presented in Section}~\ref{sec:josh_sec}, which works in terms of Eulerian variables and the conventional four-dimensional guiding center phase space. This alternative derivation shows that the new model may also be derived using an extension of the Hamiltonian method applied in \cite{Tronci2010} to derive a full-orbit PCS model.
Finally, {S}ection \ref{sec:discussion} {briefly compares} the new model to the existing PCS systems currently implemented and examines the emergence of {the cross-helicity conservation that} is lost in conventional approaches.

\section{Pressure-coupling schemes for Vlasov-MHD\label{sec:Vlasov-PCS}} 
{This section considers} the case of energetic particles described by Vlasov kinetics{---}we shall deal with the guiding center approximation in later sections. As in the standard approach, the bulk plasma is described by ideal MHD (since here we are interested in the  case of conservative dynamics) {so we} dub the overall hybrid kinetic-fluid model {\it Vlasov-MHD}.

\subsection{The CCS and non-conservative PCS models\label{non-cons}}
The standard PCS appearing in the literature \cite{Cheng,FuPark,KimEtAl,ParkEtAl} can be easily formulated starting from the following equations for the CCS \cite{BeGoFrTrCr,BelovaPark}, which {serve as} our point of departure in this discussion:
\begin{align}\label{cc-hybrid-momentum}
&\rho\partial_t\bu+\rho\left(\bu\cdot\nabla\right)\bu = \left(q_hn_h\bu -a_h\bK_h+\mathbf{J}\right)\times\bB
-\nabla\mathsf{p}\,,
\\
& \partial_t\rho+\nabla\cdot\left(\rho\bu\right)=0\,,
\label{cc-hybrid-dens}
\\
& \partial_t f+\bv\cdot\nabla f+a_h\left(\bv-\bu\right)\times\bB\cdot\frac{\pa f}{\pa\bv}=0\,, \label{cc-hybrid-Vlasov}
\\
& \partial_t\bB=\nabla\times\left(\bu\times\bB\right) \label{cc-hybrid-end} \,.
\end{align}
Here $\bu$ is the bulk fluid velocity, $\bJ=\mu_0^{-1}\nabla\times\bB$ is the usual MHD current, $q_h$ and $m_h$ are the charge and the mass of the energetic particles respectively, and $a_h=q_h/m_h$. In addition, {the energetic particle density on phase space, $f(\bx,\bv)$, generates} the following moments
\[
n_h=\int\!f\,\de^3v
\,,\qquad
{\bf K}_{h}=m_h\int\!\mathbf{v}\,f\,\de^3v
\,.
\]
The interested reader can find the detailed derivation of the equations \eqref{cc-hybrid-momentum}-\eqref{cc-hybrid-end} in both the original works \cite{ParkEtAl} as well as in more recent papers \cite{BuTr,TrTaCaMo}.
Once the CCS is obtained, the pressure-coupling scheme is conventionally derived by considering the evolution of the total momentum
\begin{equation}\label{totmom}
\bM:=\rho\bu+\bK_h
\,,
\end{equation}
{drawing on} the moment equation
\beq
\frac{\partial \bK_h}{\partial t}=-\nabla\cdot\Bbb{P}_h+a_h(\bK_h-m_hn_h\bu)\times\bB\,,
\label{Vlasov-K}
\eeq
along with  the definition of the pressure tensor
\begin{equation}
\Bbb{P}_h=m_h\int\!\bv\bv\,f(\bx,\bv)\,\de^3v\,.
\label{presstensor}
\end{equation}
As a result, we have (cf.\ Eq.~(1) of \cite{ParkEtAl})
\begin{equation}\label{primitive}
\frac{\partial\bK_h}{\partial t}+\rho\left(\frac{\partial\bu}{\partial t}+\bu\cdot\nabla\bu\right)=-\nabla\cdot\Bbb{P}_h-\nabla\mathsf{p}+ \bJ\times \bB\,.
\end{equation}
At this point, it is assumed that the population of energetic particles is rarefied enough that its momentum contribution in the fluid momentum equation~\eqref{primitive} can be neglected. Then, {by} neglecting the first term in the LHS of \eqref{primitive}, we obtain the system:  
\begin{align}
&\rho\left(\frac{\partial\bU}{\partial t} +\bU\cdot\nabla\bU\right)=-\nabla{\sf p}-\nabla\cdot\Bbb{P}_h
+ \bJ \times \bB\,,
\label{PCS-FuPark}
\\
& \frac{\partial\rho}{\partial t} + \nabla\cdot\left(\rho\bU\right)=0 \,,
\label{pc-hybrid-mass}
\\
&
\frac{\partial f}{\partial t}+\bv\cdot\frac{\partial f}{\partial \bx}+a_h\left(\bv-\bU\right)\times\bB\cdot\frac{\partial f}{\partial \bv}=0\,,
\label{kinetic-PCS}
\\
&
\frac{\partial\bB}{\partial t}=\nabla\times\left(\bU\times\bB\right)
\label{pc-hybrid-end}
\,,
\end{align}
where we have changed the MHD velocity notation into $\bU$ to distinguish between the CCS and the PCS models.
Notice that Eq.~\eqref{PCS-FuPark} is identical to the bulk momentum equation of  the hybrid PCS of Fu and Park (see equations (1) in \cite{FuPark,FuParketAl}), which also includes \eqref{pc-hybrid-mass} and \eqref{pc-hybrid-end} (while replacing Vlasov dynamics by its gyrokinetic approximation). 
 Analogous PCS models with the same fluid equation \eqref{PCS-FuPark} have been formulated by Cheng \cite{Cheng} (see equation (1) therein) and Park \emph{et al.}\  \cite{ParkEtAl} (see equation (3) therein). In some situation{s t}he tensor \eqref{presstensor} in \eqref{PCS-FuPark} is replaced by the relative pressure tensor $\widetilde{\Bbb{P}}_h=m_h\int\left(\bv-\boldsymbol{V}\right)\left(\bv-\boldsymbol{V}\right)f\,\de^3v$ (e.g.,  the PCS model proposed by Kim, Sovinec and Parker \cite{Kim,KimEtAl,TaBrKi}), where $\boldsymbol{V}={\bf K}_{h}/n_h$.

As discussed in \cite{TrTaCaMo}, equations \eqref{PCS-FuPark}-\eqref{pc-hybrid-end} do not conserve energy. When their incompressible limit is linearized around a static isotropic equilibrium, plane Alfv\`en waves become unstable at frequencies above the cyclotron frequency. This inconsistency can {actually be} removed by adding specific force terms that are required to ensure energy conservation. These force terms were first found in \cite{Tronci2010} by applying Hamiltonian methods and were later cast in the variational framework \cite{HoTr2011}. Before moving on in this direction, we discuss an important point about the above derivation of the equations  \eqref{PCS-FuPark}-\eqref{pc-hybrid-end}. Specifically, we notice that the approximation $\partial\bK_h/\partial t\simeq0$ of rarefied energetic component  was performed on the equation \eqref{primitive} for the total momentum \eqref{totmom}, which in turn involves a change of frame. Thus, an approximation was performed on a fluid equation corresponding to a new frame moving with the velocity $\bM/\rho$. Since this frame is a non-inertial frame, we would expect inertial force terms to appear in the equations of motion so that eventually the energy balance can be preserved. However, such force terms do not emerge when the approximation is performed by simply neglecting terms in the equations and other techniques need to be exploited in order to keep track of the relative motion \cite{ThMc}.

{Hamiltonian methods were used in} \cite{Tronci2010} so that the approximation of a negligible energetic particle momentum was performed by neglecting the corresponding momentum terms in the expression of the total energy (i.e. the Hamiltonian), {\it after} {expressing the energy} in terms of the total momentum $\bM$. Depending on the terms that are neglected {in the energy}, we are led to conservative PCS models involving the standard pressure tensor $\Bbb{P}_h$ or the relative pressure $\widetilde{\Bbb{P}}_h$ mentioned earlier~\cite{Tronci2010}. {The accompanying kinetic equation in the latter case} becomes {so cumbersome that} the first variant (involving $\Bbb{P}_h$) was {naturally} given more attention.

The conservative (Hamiltonian) variant of the PCS in \eqref{PCS-FuPark}-\eqref{pc-hybrid-end} reads as follows:
\begin{align}\label{PCS-MHD1}
&\rho\frac{\partial \bU}{\partial
t}+\rho(\bU\cdot\nabla)\bU=-\nabla{\sf p}-\nabla\cdot\Bbb{P}_h +\bJ\times \bB\,,
\\\label{PCS-MHD2}
&\frac{\partial f}{\partial
t}+\left(\bU+\bv\right)\cdot\nabla f+\big[a_h\bE-\nabla\bU\cdot\bv+a_h({\bv}+\bU)\times
\bB)\big]\cdot\frac{\partial f}{\partial\bv}=0\,,
\\\label{PCS-MHD3}
&\frac{\partial \rho}{\partial t}+\nabla\cdot(\rho\,\bU)=0
\,,\qquad \frac{\partial \bB}{\partial
t}=\nabla\times\left(\bU\times\bB\right) \,.
\end{align}
While the fluid equation \eqref{PCS-MHD1} is identical to \eqref{PCS-FuPark},  the kinetic equation \eqref{PCS-MHD2} differs substantially from \eqref{kinetic-PCS} since particles are now moving with the velocity $\bv+\bU$,  {indicating a frame change}. We notice that this frame change is accompanied by the inertial force $-m_h\nabla\bU\cdot\bv$ thereby leading to the effective electric field $\bE_\textit{\tiny eff}=\bE-a_h^{-1}\nabla(\bU\cdot\bv)$. Here we recall the ideal Ohm's law $\bE=-\bU\times\bB$. 

As anticipated above, {the system} \eqref{PCS-MHD1}-\eqref{PCS-MHD3} {in} \cite{Tronci2010} is derived by first rewriting the Hamiltonian in the frame moving with velocity $\bM/\rho$ and then by neglecting $\bK_h$-contributions in the expression of the energy, which then reads exactly as the energy conserved by the CCS \eqref{cc-hybrid-momentum}-\eqref{cc-hybrid-end}, that is
\[
E=\frac1{2}\int\rho|\bU|^2\,\de^3x+\int\rho\,\mathcal{U}(\rho)\,\de^3x+\frac{m_h}2\iint f |\bv|^2\,\de^3x\,\de^3v
+\frac1{2\mu_0}\int|\bB|^2\,\de^3x
\,.
\]
{The MHD bulk internal energy is denoted here by $\mathcal{U}$,} and in this case we limit to the barotropic closure $\mathcal{U}=\mathcal{U}(\rho)$ for simplicity. 
Some of the physical features of the system \eqref{PCS-MHD1}-\eqref{PCS-MHD3} have been studied in \cite{TrTaCaMo}, where it was shown that kinetic damping is consistently recovered, precisely in the frequency region where the non-conservative model \eqref{PCS-FuPark}-\eqref{pc-hybrid-end} exhibit spurious instabilities.

\subsection{Variational setting:  Lagrangian vs. Eulerian variables\label{sec:var1}}

The frame change underlying the PCS reflects an intricate variational structure \cite{HoTr2011}, which we are now going to summarize. Since in equation \eqref{PCS-MHD2} the energetic particles move relative to the fluid frame, this implies a specific decomposition of the Lagrangian particle paths as follows: upon denoting the paths by $\boldsymbol{z}(\bz_0)=(\boldsymbol{q}(\bq_0,\bv_0),\boldsymbol{v}(\bq_0,\bv_0))$, we write
\beq\label{paths}
\boldsymbol{z}(\bz_0)=\mathcal{T}\!\boldsymbol{\eta}\big(\boldsymbol{\psi}(\bz_0)\big)\,,
\eeq
where both  $\boldsymbol{\psi}$ and  $\mathcal{T}\!\boldsymbol{\eta}$ are smooth invertible paths on phase space. However, while $\boldsymbol{\psi}$ is a path of generic type, $\mathcal{T}\!\boldsymbol{\eta}$ is actually induced by the MHD fluid paths {moving with Lagrangian velocity $\dot{\boldsymbol{\eta}}(\bx_0)=\bU(\boldsymbol{\eta}(\bx_0))$: the phase}-space path $\mathcal{T}\!\boldsymbol{\eta}$ is the {\it tangent lift} of the configuration path $\boldsymbol{\eta}$. Explicit formulas for tangent lifts are found in standard textbooks such as \cite{MaRa,HoScSt}, and in this case we have
\[
\mathcal{T}\!\boldsymbol{\eta}(\bx_0,\bv_0)=\Big(\boldsymbol{\eta}(\bx_0),\,\bv_0\cdot\nabla\boldsymbol{\eta}(\bx_0)\Big).
\]
Generally speaking, any fluid configuration path $\boldsymbol{\eta}$ generates a phase space path given by $\mathcal{T}\!\boldsymbol{\eta}$. According to formula \eqref{paths}, the particles are first pushed by their own phase space flow $\boldsymbol\psi$ and then are taken into the fluid frame by the tangent lift $\mathcal{T}\!\boldsymbol{\eta}$. 

It is a vector calculus exercise to verify that the time derivative of \eqref{paths} yields
\begin{equation}\label{Zdot}
	\dot{\boldsymbol{z}}=\bcalX_\bU(\boldsymbol{z})+\bcalX(\boldsymbol{z})\,,
\end{equation}
where $\bcalX_\bU(\bx,\bv)=(\bU(\bx),\bv\cdot\nabla\bU(\bx))$ and 
\beq\label{sdp-vf}
\bcalX(\bx,\bv)=\left[\dot{\boldsymbol{\psi}}(\boldsymbol{\psi}^{-1}(\mathcal{T}\!\boldsymbol\eta^{-1}(\bx,\bv)))\right]\cdot\nabla\mathcal{T}\!\boldsymbol\eta(\bx,\bv)
=:
\Big(\bw(\bx,\bv),\ba(\bx,\bv)\Big)
\,.
\eeq
We denote the spatial and velocity components of the phase space vector field $\bcalX$ respectively by $(\bw,\ba)$, while $\nabla\mathcal{T}\!\boldsymbol\eta$ denotes the Jacobian matrix of the tangent lift $\mathcal{T}\!\boldsymbol\eta$. In standard differential geometry \cite{MaRa,HoScSt}, the relation \eqref{sdp-vf} can be written in terms of the {pushforward $\mathcal{T}\!\boldsymbol\eta_*$ as $\bcalX=\mathcal{T}\!\boldsymbol\eta_*[\dot{\boldsymbol{\psi}}(\boldsymbol{\psi}^{-1})]$.}

The intricate variational structure of the conservative PCS is inherited by the strongly coupled kinematics of the Lagrangian fluid and particle paths. In Euler-Poincar\'e theory \cite{HoMaRa1998} this is known as semidirect-product group structure and was recently exploite{d i}n fully kinetic models {in order to split} the mean velocity $\bV=\int \!\bv\,f\,\de^3v\,/\int\! f\,\de^3v$ and its fluctuations $\bc=\bv-\bV$ \cite{Tronci2014}. The discussion in this section {closely follows} the treatment in \cite{HoTr2011}, where all relations are presented in more detail. In \cite{HoTr2011}, a slight variant of the following PCS Lagrangian was presented:
\begin{multline}
l(\bU,\rho,\bcalX,f,\bA)=\frac1{2}\int\!
\rho\,{\left|\bU\right|^2}\,\de^3x -\int\!\rho\,
\mathcal{U}(\rho)\,\de^3x-\frac1{2\mu_0}\int
\left|\nabla\times\bA\right|^2\de^3x
\\
+\int \!f\left[\left(m_h\bv+q_h\bA
\right)\cdot\bw-\frac{m_h}2 |\bv|^2\right]\de^3x\,\de^3v
\,.
 \label{PCS-Lagrangian}
\end{multline}
This {simply amounts} to the sum of Newcomb's MHD Lagrangian \cite{Newcomb} and the (Eulerian) phase space Lagrangian for the energetic particles, {adopted here} instead of its modified version in \cite{CeMaHo, HoTr2011} (whose variant in terms of Lagrangian paths first appeared in \cite{Low} and {which later found use} in \cite{Dewar}). Indeed in \cite{CaTr,SqQiTa,Tronci2014}, it was shown that two slightly different Lagrangians lead to the same final form of the resulting kinetic equation.
{The objects} $\rho$, $f$, and $\bA$ are frozen-in parameters, {which is to say} they satisfy (recall equation \eqref{paths})
\beq\label{pullbacks}
\rho(\boldsymbol{\eta},t)\,\de^3 {\eta} = \rho_0(\bx_0)\,\de^3 {x}_0
\,,\qquad\ 
f(\boldsymbol{z},t)\,\de^6 {z} = f_0(\z_0)\,\de^6 {z}_0
\,,\qquad\ 
\bA(\boldsymbol{\eta},t)\cdot\de\boldsymbol{\eta}=\bA_0(\bx_0)\cdot\de \bx_0
\,,
\eeq
on, by taking time derivatives,
\beq\label{var1-A}
\frac{\partial\rho}{\partial t}=-\nabla\cdot(\rho\bU)
\,,\,\qquad
\frac{\partial f}{\partial t}=-\nabla_\bz\cdot(f\bcalX+f\bcalX_\bU)
\,,\,\qquad
\frac{\partial\bA}{\partial t}=\bU\times\bB-\nabla(\bU\cdot\bA)
\,.
\eeq
In turn, the above relations produce the variations
\beq\label{var1}
\delta\rho
=-\nabla\cdot({\boldsymbol\xi}\rho)
\,,\,\qquad
 \delta f=-\nabla_\bz\cdot({\boldsymbol\Xi f+\bcalX_{\boldsymbol\xi}}f)
\,,\,\qquad
\delta \bA
=\bxi\times\bB-\nabla(\bxi\cdot\bA)
\,,
\eeq
where $\bxi$ is an arbitrary vector field such that $\delta\boldsymbol\eta(\bx_0)=\bxi(\boldsymbol\eta(\bx_0))$, the vector field $\bcalX_{\boldsymbol\xi}$ is given by $\bcalX_{\boldsymbol\xi}=(\bxi,\bv\cdot\nabla\bxi)$, and $\boldsymbol\Xi$ is another arbitrary vector field on phase space that relates to $\delta \boldsymbol\psi$ in the same way that  the Eulerian vector field $\bcalX$ relates to the time derivative $\dot{\boldsymbol\psi}$ in  \eqref{sdp-vf} \cite{HoTr2011}. Analogous formulas hold for the variations of $\bU$ and $\bcalX${,}
\beq\label{var2}
\delta\bU=\partial_t\bxi+[\bU,\bxi]
\,,\qquad\quad
\delta \bcalX = \partial_t \boldsymbol\Xi +[\bcalX,\boldsymbol\Xi] + [\bcalX_\bU,\boldsymbol\Xi]-[\bcalX_\bxi,\bcalX]
\,,
\eeq
where we have used the notation $[\boldsymbol{P},\boldsymbol{R}]=(\boldsymbol{P}\cdot\nabla\boldsymbol{R})-(\boldsymbol{R}\cdot\nabla)\boldsymbol{P}$ for the vector field commutator ({g}radients act on the same space where $\boldsymbol{P}$ and $\boldsymbol{R}$ are defined). All these formulas are {features of the} background material already discussed extensively in \cite{HoTr2011,Tronci2014} and one can verify that using \eqref{var1} and \eqref{var2} in Hamilton's principle,
\beq\label{HP}
\delta\int_{t_1}^{t_2} l(\bU,\rho,\bA,\bcalX,f)\,\de t=0\,,
\eeq
leads to the conservative PCS equations \eqref{PCS-MHD1}-\eqref{PCS-MHD3}. Besides energy conservation, it is important to notice that these equations preserve a  hybrid expression of the cross-helicity as follows \cite{HoTr2011}:
\beq
\frac{\de}{\de t}\int \bB\cdot(\bU-\rho^{-1}\bK_h)\,\de^3x = 0
\label{CH}
\eeq
This conservation law has enabled a thorough Lyapunov stability study \cite{MoTaTr2014-1,MoTaTr2014}.

It is the purpose of this paper to extend this construction to the case of energetic particles undergoing guiding center motion. However, before embarking on this task, it will be useful to take a closer look at the general structure of the Euler-Poincar\'e equations of the PCS.

\subsection{General form of the Euler-Poincar\'e equations\label{sec:EP}}
{We will use this section to present} the general form of the Euler-Poincar\'e equations of the PCS. To begin, we consider an arbitrary Lagrangian $l$ depending on the variables
\[
(\bU,\rho,\bA,\bcalX,f)\,,
\]
as they are defined in the previous section. As shown in \cite{HoTr2011}, the general form of the Euler-Poincar\'e equations produced by the variational principle \eqref{HP} is as follows.
\begin{align}\nonumber
&\frac{\partial}{\partial t}\frac{\delta l}{\delta
{\bU}}+\pounds_{\bU}\,\frac{\delta l}{\delta \bU}=
\rho\nabla\frac{\delta l}{\delta \rho}-\frac{\delta l}{\delta
\bA}\times\bB
+\left(\nabla\cdot\frac{\delta l}{\delta \bA}\right)\bA
\\
&\hspace{3.4cm} -
\int\!\left(\!\pounds_{\bcalX}\frac{\delta \ell}{\delta
\bcalX}- f\nabla_{\bz}\frac{\delta\ell}{\delta f}\right)_{\!\!\bq}
\!\de^3v
+
\nabla\cdot\!\int\!\bv\left(\!\pounds_{ \bcalX}\frac{\delta
\ell}{\delta  \bcalX}-f\,\nabla_{\bz}\frac{\delta\ell}{\delta f}\right)_{\!\!\bv}
\de^3v\,,
\label{EP-PCS1}
\\\label{EP-PCS3}
&\frac{\partial \rho}{\partial t}+\nabla\cdot( \rho{\bU})=\,0\,, \quad\
\frac{\partial \bA}{\partial t}+\nabla(\bU\cdot \bA)=\bU\times\bB\,,
\\\label{EP-PCS2}
&\frac{\partial}{\partial t}\frac{\delta l}{\delta
\bcalX}+\pounds_{\bcalX+\bcalX_\bU}\,\frac{\delta l}{\delta \bcalX}=
f\,\nabla_{\bz}\frac{\delta l}{\delta f}\,,
\\\label{EP-PCS4}
&\frac{\partial f}{\partial t}+\nabla_\bz\cdot({f\bcalX+f\bcalX_\bU})=0\,.
\end{align}
The subscripts $\bx$ and $\bv$ denote respectively the spatial and velocity components of vector functions on phase space. For example,  $(\delta l/\delta \bcalX)_\bx=\delta l/\delta \bu$ and  $(\delta l/\delta \bcalX)_\bv=\delta l/\delta \ba$. We have used the following definition of functional derivative:
\[
\delta \mathcal{F}(\boldsymbol\phi):=\int_{\mathcal{D}}\frac{\delta\mathcal{F}}{\delta \boldsymbol\phi}\cdot\delta\boldsymbol\phi
\,,
\]
where $\mathcal{F}$ is any functional of some function (possibly, a vector function) $\boldsymbol\phi$ on some domain $\mathcal{D}$ in either the configuration space or  the phase space. {We have also} made use of the Lie derivative operator \cite{HoMaRa1998,IlLa},
\[
\pounds_{\bY}\frac{\delta\mathcal{F}}{\delta \bY}=(\bY\cdot\nabla)\frac{\delta\mathcal{F}}{\delta \bY}
+(\nabla\cdot\bY)\frac{\delta\mathcal{F}}{\delta \bY}
+
\nabla\bY\cdot\frac{\delta\mathcal{F}}{\delta \bY}
\,,
\]
{of a $1$-form density}, which can be defined on either the configuration space or the phase space. Here $\bY$ is some vector field and $\mathcal{F}$ is an arbitrary functional.

A remarkable property of the above general equations is as follows (see Proposition 4.4 in \cite{HoTr2011}):
\begin{multline}\label{KM-momap}
\left(\frac{\partial}{\partial t}+\pounds_{\bU}\right)\left(\frac{\delta l}{\delta \bU}-\int\!\frac{\delta l}{\delta \bw}\,\de^3v+\int\!\left(\bv\cdot\nabla\right)\frac{\delta l}{\delta \ba}\,\de^3v\right)
\\=
\rho\,\nabla\frac{\delta l}{\delta
\rho}-\frac{\delta l}{\delta
\bA}\times\bB+\left(\nabla\cdot\frac{\delta
l}{\delta \bA}\right)\bA
\,,
\end{multline}
which lies behind the preservation of the cross-helicity \eqref{CH}.

At this poin{t w}e have constructed a systematic method for constructing hybrid pressure-coupling schemes, depending on the form of the Lagrangian $l$. {We shall see that} this framework applies to energetic particles undergoing guiding center motion without substantial modifications, provided extra care is taken to deal with the geometric structure of Littlejohn's theory \cite{Littlejohn}. An alternative variational setting of hybrid plasma models for energetic particles will be also presented in Section \ref{sec:josh_sec}.

{\color{black}

\subsection{Variational relationship between CCS and PCS\label{sec:CCStoPCS}}
As explained in Section \ref{non-cons}, the non-conservative PCS is obtained from the CCS by neglecting the term $\partial\bK_h/\partial t$ in the momentum equation \eqref{primitive}, but in the energy conserving case, the PCS is obtained by the very different procedure \cite{Tronci2010} of neglecting the kinetic momentum $\bK_h$ in the expression of the conserved energy for the CCS.
However, the relationship between the two coupling schemes on the variational side has remained elusive. In this section we will shed some light on this relationship and show that a similar approximation in the momentum to the Hamiltonian derivation provides the link between the two models, but with the important realization that the approximation takes place within a shifted frame that induces a similar variational structure to that of Section \ref{sec:EP}.

As a first step, we briefly present the variational setting of the CCS \eqref{cc-hybrid-momentum}-\eqref{cc-hybrid-end}. Since there is no frame change involved in the formulation of the CCS, the dynamics of particle paths in \eqref{paths} is replaced by the simpler relation 
\beq\label{paths2}
\boldsymbol{z}(\bz_0)=\boldsymbol{\Psi}(\bz_0)\,,
\eeq
where  $\boldsymbol{\Psi}$ is a time-dependent path on phase space. Then, the second in \eqref{pullbacks} simplifies the corresponding relations in \eqref{var1-A} and \eqref{var1}, which become  
\[
\frac{\partial f}{\partial t}=-\nabla_\bz\cdot(f{\bcalX}_\textit{\tiny \!CCS} )
\,,\qquad\qquad
\delta f=-\nabla_\bz\cdot(f{\boldsymbol\Xi}_\textit{\tiny CCS} )
\,.
\]
Here, we have defined the phase space vector fields ${\bcalX}_\textit{\tiny \!CCS}$ and ${\boldsymbol\Xi}_\textit{\tiny CCS}$ in such a way that
\[
\dot{\boldsymbol{z}}={\bcalX}_\textit{\tiny \!CCS\,}(\boldsymbol{z})
\,,\qquad\qquad
\delta{\boldsymbol{z}}={\bcalXi}_\textit{\tiny CCS}(\boldsymbol{z})
\,,
\]
and thus ${\bcalX}_\textit{\tiny \!CCS\,}(\bz)=\dot{\boldsymbol{\Psi}}(\boldsymbol{\Psi}^{-1}(\bz))$ and ${\bcalXi}_\textit{\tiny CCS}(\bz)=\delta{\boldsymbol{\Psi}}(\boldsymbol{\Psi}^{-1}(\bz))$. A direct calculation shows that these vector fields satisfy the relation
\[
\delta {\bcalX}_\textit{\tiny \!CCS\,}=\partial_t{\bcalXi}_\textit{\tiny CCS}+[{\bcalX}_\textit{\tiny \!CCS\,},{\bcalXi}_\textit{\tiny CCS}]
\,,
\]
where we have used the same notation as in \eqref{var2}. 
Then, upon introducing $\bu(\bx)=\dot{\boldsymbol\eta}({\boldsymbol\eta}^{-1}(\bx))$, using the  definitions  above and the other relations in \eqref{pullbacks} for $\rho$ and $\bA$ in the following Lagrangian \cite{HoTr2011} yields the CCS \eqref{cc-hybrid-momentum}-\eqref{cc-hybrid-end}:
\begin{equation}\label{CCS-lag}
\begin{aligned}
	\ell_\textit{\tiny CCS\,}^\textrm{\tiny\,(1)}( f,{\bcalX}_\textit{\tiny \!CCS\,},\rho,\bu,\bA)=&
	\int\!\bigg[\frac1{2}
\rho\,{\left|\bu\right|^2} -\rho\,
\mathcal{U}(\rho)-\frac1{2\mu_0}
\left|\nabla\times\bA\right|^2\bigg]\de^3x
\\
	&+\int \!f\bigg[(m_h\bv+q_h\bA)\cdot\bw_\textit{\tiny CCS}
	-\frac{m_h}{2}|\bv|^2-q_h\bA\cdot\bu\bigg]\de^3 x\de^3 v
\,,
\end{aligned}
\end{equation}
where we have used the notation ${\bcalX}_\textit{\tiny \!CCS\,}=(\bw_\textit{\tiny CCS},\ba_\textit{\tiny CCS}).$
This construction was extended in \cite{BuTr} to the case of energetic particles undergoing guiding-center motion.

At this point, we present an alternative variational formulation of the same CCS  in such a way that its equations of motion \eqref{cc-hybrid-momentum}-\eqref{cc-hybrid-end} can be obtained by using precisely the same construction as in Section \ref{sec:EP}. To this purpose, 
 we consider the case where the hot particle pathways are shifted into the frame of the fluid motion by simply writing the phase space path $\boldsymbol{\Psi}$ as $\boldsymbol{\Psi}=\mathcal{T}\!\boldsymbol{\eta}\circ\boldsymbol{\psi}$ (here, the symbol $\circ$ denotes the usual  composition of functions) or, equivalently, 
\beq\label{pathdecomp}
\boldsymbol{\Psi}(\bz_0)=\mathcal{T}\!\boldsymbol{\eta}\big(\boldsymbol{\psi}(\bz_0)\big)
\eeq
This decomposition of the phase space path affects the expression of the particle trajectories in such a way that \eqref{paths2} is now replaced by its previous version \eqref{paths}, thereby returning $\dot{\boldsymbol{z}}=\bcalX_\bu(\boldsymbol{z})+\bcalX(\boldsymbol{z})$ as in \eqref{Zdot}. In turn, this implies the relations
\[
{\bcalX}_\textit{\tiny \!CCS\,}=\bcalX+\bcalX_\bu
\,,\qquad\qquad
{\bcalXi}_\textit{\tiny CCS}={\bcalXi}+\bcalX_\bxi
\,,
\]
where we have used again the same notation as in Section \ref{sec:var1}. Therefore, replacing the relations above in the Lagrangian \eqref{CCS-lag} yields its alternative formulation
\begin{equation}\label{CCS-lag2}
\begin{aligned}
	\ell_\textit{\tiny CCS\,}^\textrm{\tiny\,(2)}( f,{\bcalX},\rho,\bu,\bA)=&
	\int\!\bigg[\frac1{2}
\rho\,{\left|\bu\right|^2} -\rho\,
\mathcal{U}(\rho)-\frac1{2\mu_0}
\left|\nabla\times\bA\right|^2\bigg]\de^3x
\\
	&+\int \!f\bigg[(m_h\bv+q_h\bA)\cdot\bw
	-\frac{m_h}{2}|\bv|^2+m_h\bv\cdot\bu\bigg]\de^3 x\de^3 v
\,.
\end{aligned}
\end{equation}
A direct verification shows that using the Lagrangian $\ell_\textit{\tiny CCS\,}^\textrm{\tiny\,(2)}$ in the equations \eqref{EP-PCS1}-\eqref{EP-PCS4} consistently reproduces the dynamical equations \eqref{cc-hybrid-momentum}-\eqref{cc-hybrid-end} of the full-orbit CCS model.

We observe that the last term in the Lagrangian \eqref{CCS-lag2} can be rewritten as
\[
m_h\int \!f \bv\cdot\bu\,\de^3 x\de^3 v=m_h\int \! \bK_h\cdot\bu\,\de^3 x
\,.
\]
Thus, if we adopt the standard assumption that the averaged kinetic momentum $\bK_h$ produces negligible effects, the approximation $\bK_h=m_h\int \!f\bv\,\de v\simeq0$ is applied to the Lagrangian~\eqref{CCS-lag2}, thereby producing the Eulerian PCS Lagrangian \eqref{PCS-Lagrangian} and returning the conservative PCS model~\eqref{EP-PCS1}-\eqref{EP-PCS4} upon replacing $\bu$ by $\bU$. The relationship between the CCS and PCS on the variational side, then, is through an approximation in a shifted frame.

This section addressed an important question and unfolded the intricate relationship between the variational structures of the CCS and the conservative PCS. As we showed, the latter is obtained from the first by simply neglecting momentum contributions of the energetic particles. The explicit particle path decomposition \eqref{pathdecomp} sheds a new light on the kinematics of the conservative PCS, which is now fully explained in terms of Lagrangian trajectories.

}

\section{Hybrid PCS in the guiding center approximation\label{sec:2}}
This section applies the Euler-Poincar\'e framework summarized in the previous sections to the case of hybrid PCS models where energetic particles are described in terms of guiding center theory. After some general considerations we shall move on to apply the mathematical framework.

\subsection{Non-conservative PCS models from the CCS\label{sec:NC-PCS}} 
{Let us} formulate a PCS model by mimicking the procedure leading to the non-conservative model \eqref{PCS-FuPark}-\eqref{pc-hybrid-end}. {We will see that although it destroys} the energy balance, this procedure {illuminates} some of the features that will emerge in the conservative case.

We start with the current-coupling scheme. As mentioned in the introduction, the only available code using the CCS in the guiding center approximation is the MEGA code \cite{Todo,ToSaWaWaHo}, {despite being} based on continuum equations lacking the correct energy balance. A conservative variant of the equations underlying the MEGA code has recently been proposed in \cite{BuTr} and reads as follows:
\begin{align}\label{cc-hybrid-momentum-GC}
&\rho\partial_t\bu+\rho\left(\bu\cdot\nabla\right)\bu
=\Big(\bJ+q_hn_h\bu-\bJ_{\rm gc}-\nabla\times{\bf M}_{\rm gc}\Big)\times\bB-\nabla{\sf p}\,,
\\
& \partial_t\rho+\nabla\cdot\left(\rho\bu\right)=0 \label{cc-hybrid-mass}\,,
\\
&\label{cc-hybrid-mass-DK}
\partial_t f+\nabla\cdot\left[\frac{1}{B^*_\|}\Big(v_\|\bB^*-\bb\times\bE^*\Big)f\right]+\partial_{v_\|\!}\left[\frac{{a_h}}{B^*_\|}\left(\bB^*\cdot\bE^*\right) f\right]=0\,,
\\
&
\partial_t\bB=\nabla\times\left(\bu\times\bB\right)
\label{cc-hybrid-end-GC}
\,,
\end{align}
where we {have} used the standard notation
\beq
\bB^*(\bx,t):=\bB(\bx,t)+a_h^{-1}v_\parallel \nabla\times\bb(\bx,t)
\,,\qquad\ 
\bE^*(\bx,t):=\bE(\bx,t)-a_h^{-1}v_\parallel \partial_t\bb(\bx,t)-{q_h^{-1}}\mu \nabla B(\bx,t)\,,
\label{EffFields}
\eeq
and we have introduced the guiding center current
\beq
bJ_{\rm gc}(\bx,t)=\int_{\!\mu\,}\!\frac{q_h}{B^*_\| (\bx,t)}\Big(v_\|\bB^*(\bx,t)-\bb(\bx,t)\times\bE^*(\bx,t)\Big)f(\bx,\bv,t)\,\de v_\|\,,
\label{J+M}
\eeq
as well as the guiding center magnetization
\begin{align}
{\bf M}_{\rm gc}(\bx,t)
=-\int_{\!\mu\,}\!\bigg[\mu \bb(\bx,t)-\frac{m_hv_\|}{B(\bx,t)B^*_\|(\bx,t)}\Big(v_\|\bB^*_\perp(\bx,t)-\bb(\bx,t)\times\bE^*(\bx,t)\Big)\bigg]f(\bx,\bv,t)\,\de v_\|
\,.
\label{magn-corr}
\end{align}
While we have included the explicit dependence of the fields above, this will be omitted going forward for ease of notation.
The symbol $\int_\mu$ stands for $\iint\!\de\mu$, where the magnetic moment coordinate is regarded as a fixed parameter in the probability density on phase space $f$. Notice that in this paper we are using $f$ to denote the phase space density retaining the Liouville measure $B^*_\|$ (other works use the capital $F$ to denote the same object).

While the current expression \eqref{J+M} is well-consolidated in the theory, the inner parenthesis appearing in the magnetization \eqref{magn-corr} deserves some attention. Th{e} term is a moving dipole contribution \cite{Ka} that is needed in the theory to ensure the correct energy and momentum balance. This term emerges naturally within the variational framework, as shown in \cite{Evstatiev_2014,BrTr}{, and w}e remark that {it is} absent in the equations underlying the MEGA code \cite{ToSa,ToSaWaWaHo}, which in turn identify $B\simeq B^*_\|$, thereby leading to further cancelations in the momentum equation \eqref{cc-hybrid-momentum-GC}.

At this point, we follow the procedure outlined in {S}ection \ref{non-cons} by considering the total momentum
\begin{align}
\bM=\rho\bu+\bK_{\rm gc}\,,\label{eq:total_gc_momentum}
\end{align}
where we have denoted 
\begin{align}
\bK_{\rm gc}=m_h\int_\mu v_\|\,f\bb\,\de v_\|\label{eq:hot_gc_momentum}
\,.
\end{align}
To this purpose, we take the first moment of equation \eqref{cc-hybrid-mass-DK} to write
\beq\label{gc-K}
\frac{\partial\bK_{\rm gc}}{\partial t}=
-
\nabla\cdot\Bbb{P}_{\rm gc}
+\big(\bJ_{\rm gc}+\nabla\times{\bf M}_{\rm gc}-q_hn_h\bu\big)\times\bB
\,,
\eeq
where
\beq
\label{gc-P}
\Bbb{P}_{\rm gc}=
\int_\mu \!\left({m_h}v_\|^2\bb\bb+\mu B(\boldsymbol{1}-\bb\bb)
+
{m_h}v_\|\boldsymbol{\sf w}_\perp\bb+{m_h}v_\|\bb\boldsymbol{\sf w}_\perp
\right)f\,
\de v_\|\,,
\eeq
and $\boldsymbol{\sf w}=(v_\|\bB^*-\bb\times\bE^*)/B^*_\|$, so that $\boldsymbol{\sf w}_\perp=\boldsymbol{\sf w}-v_\|\bb$. Notice that equation \eqref{gc-K} ({whose proof is left to} Appendix \ref{app1}) is the guiding center analogue of the corresponding Vlasov moment equation \eqref{Vlasov-K}. {Meanwhile}, the guiding center pressure tensor \eqref{gc-P} coincides with the corresponding expression recently found in \cite{BrTr}: when the self-consistent evolution of the electromagnetic field is considered in guiding center theory, the usual CGL stress tensor is corrected by the extra terms appearing in \eqref{gc-P}. Again, these terms are necessary to ensure the correct energy and momentum balance.

If w{e k}eep following the procedure outlined in {S}ection \ref{non-cons}, we  take the sum of the momentum equations \eqref{cc-hybrid-momentum-GC} and \eqref{gc-K} to write
\[
\frac{\partial\bK_{\rm gc}}{\partial t}+\rho\left(\frac{\partial\bu}{\partial t}+\bu\cdot\nabla\bu\right)
=
-\nabla{\sf p}-\nabla\cdot\Bbb{P}_{\rm gc} +\bJ\times \bB
\,.
\]
Then, assuming a rarefied energetic component, a non-conservative PCS in the guiding center approximation is obtained by dropping the term ${\partial\bK_{\rm gc}}/{\partial t}$ and by coupling the equation above to the equations \eqref{cc-hybrid-mass-DK}-\eqref{EffFields}. Such a non-conservative PCS model would suffer from the same drawbacks as its corresponding Vlasov model \eqref{PCS-FuPark}-\eqref{pc-hybrid-end}: unphysical energy sources may drive unphysical instabilities which could be filtered by the addition of extra dissipative terms.
In the next section we will instead derive a conservative variant of the PCS in the guiding center approximation by following the variational approach outlined in {S}ection \ref{sec:var1}.

\subsection{Formulation of the new PCS model\label{sec:newmod}}

{We will now} apply the procedure outlined in {S}ection \ref{sec:var1} to the case of energetic particles undergoing guiding center motion. We remark that this procedure cannot be applied {\em directly} for the following reason: in {S}ection \ref{sec:var1} we composed the tangent lift of fluid paths with phase space paths to obtain the overall trajectories of the energetic particles. This presented no difficulty since both tangent lifts and Vlasov phase space paths evolve in the 6-dimensional phase space. {But} in the case of guiding center theory this is no longer true, since guiding center particle paths evolve in the reduced 4-dimensional phase space, while tangent lifts still act in 6 dimensions. In order to proceed, we need to embed the guiding center phase space within the full 6-dimensional phase space. This operation introduces a redundancy which {will eventually be eliminated} by integrating out the perpendicular velocity coordinate.

{Throughout this} section, we shall write $v_\|=\bb\cdot\bv$, so that the guiding center Lagrangian \cite{CaBr,Littlejohn,Tronci2016} can be written as
\[
\ell(\boldsymbol{\sf X},\bv)=(m_h\bb\bb\cdot\bv+q_h\bA)\cdot\dot{\boldsymbol{\sf X}}-\frac{m_h}2(\bb\cdot\bv)^2-\mu B-\Phi
\,,
\]
where we have included the electrostatic potential $\Phi$ for completeness.

{As usual, $\mu$ is considered} a constant parameter to avoid carrying the gyrophase within the set of dynamical variables. {I}n order to specialize the PCS Lagrangian \eqref{PCS-Lagrangian} to the case of energetic particles undergoing guiding center motion, we modify the kinetic part of \eqref{PCS-Lagrangian} to write the following Lagrangian for the PCS in the guiding center approximation:
\begin{multline}
l(\bU,\rho,\bcalX,\hat{f},\bA)=\frac1{2}\int\!
\rho\,{\left|\bU\right|^2}\,\de^3x -\int\!\rho\,
\mathcal{U}(\rho)\,\de^3x-\frac1{2\mu_0}\int
\left|\nabla\times\bA\right|^2\de^3x
\\
+\int \!\hat{f}\left[\left(m_h\bb\bb\cdot\bv+q_h\bA
\right)\cdot{\bw}-\frac{m_h}2(\bb\cdot\bv)^2-\mu B\right]\de^3x\,\de^3v
\,.
 \label{GCPCS-Lagrangian}
\end{multline}
We emphasize that all dynamical variables are constructed exactly in the same way as in {S}ection \ref{sec:var1} and the only changes in the construction reside in the kinetic part of the Lagrangian. Specifically, $\hat{f}(\bx,\bv)$ is the phase space density in $6D$ (as opposed to $f(\bx,v_\|)$ in \eqref{cc-hybrid-mass-DK}) and we recall that $\bcalX(\bx,\bv)=(\bw(\bx,\bv),\ba(\bx,\bv))$, where we have replaced $\boldsymbol{\sf X}$ by $\bx$ to avoid unnecessary proliferation of notation.

At this point, we can simply apply the general Euler-Poincar\'e equations {of S}ection \ref{sec:EP}. We start by evaluating the functional derivatives
\[
\frac{\delta l}{\delta \ba}=0
\,,\qquad\ 
\frac{\delta l}{\delta \bw}=\hat{f}\left(m_h\bb\bb\cdot\bv+q_h\bA
\right)
\,,\qquad\ 
\frac{\delta l}{\delta \hat{f}}
=
\left(m_h\bb\bb\cdot\bv+q_h\bA
\right)\cdot\bw-\frac{m_h}2(\bb\cdot\bv)^2-\mu B\,,
\]
and by considering equation \eqref{EP-PCS2} in its spatial and velocity components. Upon dividing by $\hat{f}$, the {last of these} yields
\begin{align*}
0=\frac{\pa}{\pa\bv}(\bw+\bU)\cdot\frac{\delta l}{\delta \bw}\frac1{\hat{f}}-
\frac{\pa}{\pa\bv}\frac{\delta l}{\delta \hat{f}}
=&\,
\frac{\pa\bw}{\pa\bv}\cdot
\left(m_h\bb\bb\cdot\bv+q_h\bA
\right)
\\
&\,
-
{\frac{\pa}{\pa\bv}}\left[\left(m_h\bb\bb\cdot\bv+q_h\bA
\right)\cdot\bw-\frac{m_h}2(\bb\cdot\bv)^2\right]\,,
\end{align*}
so that, after some algebra, 
\[
\bb\cdot\bw(\bx,\bv)=\bb\cdot\bv
\,.
\]
Analogously, the spatial component of \eqref{EP-PCS2} reads
\begin{multline*}
\frac{\partial}{\partial t}\!\left(\frac1{\hat{f}}\frac{\delta l}{\delta
\bw}\right)+({\bw+\bU})\cdot{\nabla}\left(\frac1{\hat{f}}\frac{\delta l}{\delta \bw}\right)+(\ba+\bv\cdot\nabla\bU)\cdot{\frac{\pa}{\pa\bv}}\left(\frac1{\hat{f}}\frac{\delta l}{\delta
\bw}\right)
\\
+{\nabla}(\bw+\bU)\cdot\frac{\delta l}{\delta \bw}\frac1{\hat{f}}
=
{\nabla}\frac{\delta l}{\delta \hat{f}}
\,,
\end{multline*}
that is,
\begin{multline*}
\frac{\partial}{\partial t}\left(\bb\bb\cdot\bv+a_h\bA
\right)
-
({\bw+\bU})\times{\nabla}\times\left(\bb\bb\cdot\bv+a_h\bA
\right)
+\bb\bb\cdot(\ba+\bv\cdot\nabla\bU)
\\
=
{\nabla}
\left[\left(\bb\bb\cdot\bv+a_h\bA
\right)\cdot\bw-\frac{1}2(\bb\cdot\bv)^2-\mu B
-
({\bw+\bU})\cdot
\left(\bb\bb\cdot\bv+a_h\bA
\right)\right],
\end{multline*}
and, after {more} algebra,
\begin{multline}\label{mario}
a_h\bE^*-\left(\bb\cdot\bv
\right)\nabla(\bb\cdot\bU)
+a_h({\bw+\bU})\times\bB^*
\\
=
\left[\bb\cdot\ba+\bv\cdot\left(\frac{\partial\bb}{\partial t}
+(\bw+\bU)\cdot\nabla\bb
+\nabla\bU\cdot\bb\right)\right]\bb
\,.
\end{multline}
{W}e recall the definition of the effective {electromagnetic fields \eqref{EffFields} along with} ideal Ohm's law $\bE=-\bU\times\bB=-\partial_t\bA-\nabla(\bU\cdot\bA)$ (where the second equality follows from \eqref{EP-PCS3}). Crossing \eqref{mario} with $\bb$, we obtain
\beq\label{u-eq}
\bw+\bU=\left[\bb\cdot(\bv+ \bU)\right]\frac{\bB^*}{B^*_\|}-\frac{\bb}{B^*_\|}\times\left[\bE^*-a_h^{-1}\left(\bb\cdot\bv
\right)\nabla(\bb\cdot\bU)\right]
\,,
\eeq
while dotting \eqref{mario} with {$\bB^*$} gives the following expression for $\bb\cdot\ba$.
\beq\label{a-eq}
\bb\cdot\ba+\bv\cdot\left(\frac{\partial\bb}{\partial t}
+(\bw+\bU)\cdot\nabla\bb
+\nabla\bU\cdot\bb\right)
=
\frac{\bB^*}{B^*_\|}\cdot\left(a_h\bE^*-v_\|\nabla U_\|\right)
\,.
\eeq
{We are now facing} the consequence of the redundancy {we have} introduced by embedding guiding center motion in the 6-dimensional phase space: no explicit expression for $\ba_\perp$ can be found so no equation for $\hat{f}$ is available. But, we know that $\hat{f}$ satisfies the second  in \eqref{pullbacks}  (Lagrange-to-Euler map) and this allows us to find an equation of motion for 
\[
f=\int_\mu\hat{f}\,\de^2v_\perp
\,.
\] 
As proved in Appendix \ref{app2}, we obtain the following kinetic equation.
\beq
\partial_t f+{\nabla}\cdot\left[\Big((v_\|+U_\|)\frac{\bB^*}{B^*_\|}-\frac{\bb}{B^*_\|}\times(\bE^*-a_h^{-1}v_\|\nabla U_\|\Big)f\right]+\partial_{v_\|\!}\left[\frac{\bB^*}{B^*_\|}\cdot\left(a_h\bE^*-v_\|\nabla U_\|\right) f\right]=0\,.
\label{lemma}
\eeq
At this point, it is clear that the inertial force terms appearing in \eqref{PCS-MHD2} for the case of full-orbit Vlasov kinetics transfer naturally through the guiding center approximation to produce the effective electric field $\bE^*-a_h^{-1}v_\|\nabla U_\|$. While this may not sound surprising, it is not completely trivial to actually derive these terms by simply operating on the dynamical equations of the Vlasov PCS \eqref{PCS-MHD1}-\eqref{PCS-MHD3}. We incorporated these terms by applying a systematic variational framework exploiting the intricate interplay between Lagrangian paths and Eulerian variables.

As a next step, we focus on the fluid momentum equation \eqref{EP-PCS1}.  As shown in Appendix \ref{app3}, by recalling \eqref{u-eq} and by denoting $\bA^{\!*}=\bA+a_h^{-1}(\bb\cdot\bv) \bb$, we evaluate
\begin{align}\nonumber
\int\!\left(\!\pounds_{\bcalX}\frac{\delta \ell}{\delta
\bcalX}- f\nabla_{\bz}\frac{\delta\ell}{\delta \hat{f}}\right)_{\!\!\bq}
\!\de^3v
&\,
-
{\nabla}\cdot\!\int\!\bv\left(\!\pounds_{ \bcalX}\frac{\delta
\ell}{\delta  \bcalX}-\hat{f}\,\nabla\frac{\delta\ell}{\delta \hat{f}}\right)_{\!\!\bv}
\de^3v
\\
&\,
=\nabla\cdot\int_\mu{q_h}\hat{f}\bw\bA^{\!*}\,\de^3v
{-}\int_\mu\hat{f}\Big[({q_h}\nabla{\bf A}
	-{m_h}({\bb\cdot \bv})\nabla\bb)\cdot\bu
	-\mu\nabla B\notag\Big]\de^3{v}
\\
&\,
=\nabla\cdot\int_\mu{q_h}\hat{f}\bw\bA^{\!*}\de^3v
	-\int_\mu{q_h}\hat{f}\,\nabla{\bf A}\cdot\bw\,\de^3v
	-\nabla{\bf B}\cdot {\bf M}_{\rm gc}
\,,
\label{Jack}
\end{align}
where we have used the following expression of the magnetization density,
\[
\bM_{\rm gc}
=-\int_{\!\mu\,}\!\bigg[\mu \bb-\frac{m_h\bb\cdot\bv}{B}\bw_\perp\bigg]\hat{f}\,\de^3 v\,.
\]
In addition, upon denoting $\bJ_{\rm gc}=\int_\mu {q_h}\bu\,\hat{f}\,\de^3 v$, we have
\[
\frac{\delta l}{\delta \bA}=\bJ_{\rm gc}+\nabla\times\bM_{\rm gc}-\mu_0^{-1}\nabla\times\bB
\,,
\]
so that 
\beq
\bB\times\frac{\delta l}{\delta
\bA}
+\left({\nabla}\cdot\frac{\delta l}{\delta \bA}\right)\bA
=
(\bJ-\bJ_{\rm gc}-\nabla\times\bM_{\rm gc})\times\bB
+\left(\nabla\cdot {\bJ}_{\rm gc}\right)\bA
\,.
\label{Luke}
\eeq
Then, upon adding the terms in \eqref{Jack} and \eqref{Luke}, and by observing that
\begin{align}\nonumber
\nabla\bB\cdot\bM_{\rm gc}+\bB\times\nabla\times\bM_{\rm gc}
=&\ 
\nabla\cdot\left[(\bB\cdot\bM_{\rm gc})\boldsymbol{1}-\bB\bM_{\rm gc}\right]
\\
=&\ 
\nabla\cdot\int_\mu\Big[\mu B(\boldsymbol{1}
	-\bb\bb)
	-{m_h}({\bf v\cdot b})\bb\bw_\perp\Big]\hat{f}\,\de^3{v}
	\,,
\label{Mike}
\end{align}
equation \eqref{EP-PCS1} leads to
\beq
\rho\left(\frac{\partial\bU}{\partial t}+\bU\cdot\nabla\bU\right)
=
-\nabla{\sf p}-m_h\nabla\cdot\Bbb{P}_{\rm gc} +\bJ\times \bB
\,.
\label{GCPCS-eq1}
\eeq
{In the above}, we have denoted
\begin{align*}
\Bbb{P}_{\rm gc}=
\int_\mu \!\Big[{m_h}(\bb\cdot\bv)^2\bb\bb+\mu B(\boldsymbol{1}-\bb\bb)
+
{m_h}(\bb\cdot\bv)\big(\bw_\perp\bb+\bb\bw_\perp\big)
\Big]\hat{f}\,
\de^3 v
\,,
\end{align*}
which reduces exactly to the expression \eqref{gc-P}, after integration over the perpendicular velocity coordinates $\bv_\perp$.

To summarize, we have obtained the following PCS equations for a hybrid kinetic-fluid model in the guiding center approximation
\begin{align}\label{GCPCS-MHD1}
&\rho\frac{\partial \bU}{\partial
t}+\rho(\bU\cdot\nabla)\bU=-\nabla{\sf p}-{\nabla}\cdot\Bbb{P}_{\rm gc} +\bJ\times \bB\,,
\\\label{GCPCS-MHD2}
&\frac{\partial \rho}{\partial t}+{\nabla\cdot}(\rho\,\bU)=0
\,,\qquad \quad\frac{\partial \bB}{\partial
t}=\nabla\times\left(\bU\times\bB\right) \,,
\\\nonumber
&\frac{\partial f}{\partial t}+{\nabla}\cdot\left[\Big((v_\|+U_\|)\frac{\bB^*}{B^*_\|}-\frac{\bb}{B^*_\|}\times(\bE^*-a_h^{-1}v_\|\nabla U_\|\Big)f\right]
\nonumber
\\
&\hspace{7cm}
+\partial_{v_\|\!}\left[\frac{\bB^*}{B^*_\|}\cdot\left(a_h\bE^*-v_\|\nabla U_\|\right) f\right]=0
\,,
\label{GCPCS-MHD3}
\end{align}
with the definitions \eqref{EffFields} and \eqref{gc-P}, and where $\bE=-\bU\times\bB$ and $\bw$ is given by \eqref{u-eq}. By construction, this system conserves energy and momentum. This system additionally allows for cross-helicity conservation, as proved in Section~\ref{sec:discussion}.

We conclude by observing that the variational relation presented in Section \ref{sec:CCStoPCS} between the CCS and the conservative PCS  naturally extends to the case of guiding-center motion by simply implementing the particle path decomposition \eqref{pathdecomp}. In the following section, we shall present one more variational approach connecting the two models.

\subsection{An alternative variational approach\label{sec:josh_sec}}

The original derivation of the conservative Vlasov PCS in \cite{Tronci2010} used an approach that is quite different from that employed in the previous subsection, which was modeled on the methods of \cite{HoTr2011}. Starting from a Hamiltonian formulation of the conservative Vlasov CCS (also described in \cite{Tronci2010}), the authors of \cite{Tronci2010} performed a change of dependent variables that exchanged the bulk fluid velocity with the total kinetic momentum density. Then the terms in the transformed Hamiltonian proportional to the momentum density carried by the kinetic population were dropped. This suggests that a similar approach might be available to derive the new PCS model in this article. 

In this subsection we will demonstrate that our new PCS model may indeed be derived using an analogue of the approach used in \cite{Tronci2010}. Moreover, in order to avoid working directly with field-theoretic Poisson brackets (which become quite cumbersome in this case), we will formulate our discussion in terms of phase space Lagrangians, which are known to provide a convenient bridge between the Hamiltonian and Lagrangian formalisms. The basic strategy will be as follows. First we will present the phase space Lagrangian underlying the (guiding center version of the) variational CCS presented in \cite{BuTr}. Then we will perform a change of variables within the phase space Lagrangian from the bulk fluid velocity $\bu$ to the total momentum variable $\mathbf{M}$ given in equation~\eqref{eq:total_gc_momentum}. Finally we will drop terms in the Hamiltonian part of the phase space Lagrangian that are proportional to the parallel momentum of the kinetic species. The resulting simplified transformed phase space Lagrangian will reproduce the conservative PCS given in equations~\eqref{GCPCS-MHD1}-\eqref{GCPCS-MHD3}. It is worth mentioning here that this alternative derivation avoids embedding the 4-dimensional guiding center phase space into the 6-dimensional Vlasov phase space; instead the 4-dimensional guiding center phase space is employed directly throughout the discussion. 

To begin, we restate the Euler-Poincar\'e formulation of the guiding center CCS given in \cite{BuTr} using a notation that is as consistent as possible with the notation used so far. The guiding center CCS Lagrangian is given by
\begin{align}
l_{\text{CCS}}(\bu,\rho,\bcalX,f,\mathbf{A})&=\frac{1}{2}\int \rho|\bu|^2\,\mathrm{d}^3x-\int\rho\,\mathcal{U}(\rho)\,\mathrm{d}^3x-\frac{1}{2\mu_0}\int|\nabla\times\mathbf{A}|^2\,\mathrm{d}^3x\nonumber\\
&+\int_\mu f\bigg\{\left[m_h v_\parallel \boldsymbol{b}+q_h \mathbf{A}\right]\cdot\bw-\left[\frac{m_h}{2}v_\parallel^2+\mu B+q_h\bu\cdot\mathbf{A}\right]\bigg\}\,\mathrm{d}^4{z}\,,
\end{align}
where the symbols $(\bu,\rho,\bcalX,f,\mathbf{A})$ now have slightly different meanings than in previous sections. The most critical difference is that the guiding center phase space consists of points $\mathbf{z}=(\mathbf{X},v_\parallel)$ in a four-dimensional, rather than six-dimensional, space. The variable $\bu$ is the velocity of the \emph{bulk} fluid, which is expressed in terms of the bulk fluid configuration map $\boldsymbol{\eta}(\mathbf{x}_0)$ according to $\dot{\boldsymbol{\eta}}(\mathbf{x}_0)=\bu(\boldsymbol{\eta}(\mathbf{x}_0))$. The variable $\bcalX=(\boldsymbol{w},a_\parallel)$ is the velocity of the kinetic species' (four-dimensional) phase space fluid, which is similarly given in terms of the phase space fluid configuration map $\boldsymbol{z}(\mathbf{z}_0)$ as $\dot{\boldsymbol{z}}(\mathbf{z}_0)=\bcalX(\boldsymbol{z}(\mathbf{z}_0))$. Finally, the variables $\rho$, $f$, and $\mathbf{A}$ are advected parameters given in terms of the fluid and phase space configuration maps, $\boldsymbol{\eta}$ and $\boldsymbol{z}$, as in equation~\eqref{pullbacks}, but with $\boldsymbol{\eta}$ and $\boldsymbol{z}$ interpreted as just described, and $f$ is defined as in \eqref{cc-hybrid-mass-DK}. Accordingly, the constrained variations associated with the advection relations \eqref{pullbacks} are modified according to
\beq\label{ccs_var1}
\delta\rho
=-\nabla\cdot({\boldsymbol\xi}\rho)
\,,\,\qquad
 \delta f=-\nabla_\bz\cdot({\boldsymbol\Xi f})
\,,\,\qquad
\delta \bA
=\bxi\times\bB-\nabla(\bxi\cdot\bA)
\,,
\eeq
and
\beq\label{ccs_var2}
\delta\bu=\partial_t\bxi+[\bu,\bxi]
\,,\qquad\quad
\delta \bcalX = \partial_t \boldsymbol\Xi +[\bcalX,\boldsymbol\Xi], 
\,
\eeq
where $\bxi$ and $\boldsymbol{\Xi}$ are arbitrary vector fields on configuration space and four-dimensional phase space, respectively. The conservative guiding center CCS model then follows from the variational principle
\begin{align}\label{CCS_EP_VP}
\delta\int_{t_1}^{t_2}l_{\text{CCS}}(\bu,\rho,\bcalX,f,\mathbf{A})\,\mathrm{d}t=0,
\end{align}
which was shown in \cite{BuTr} to be equivalent to the system of equations \eqref{cc-hybrid-momentum-GC}-\eqref{magn-corr}.

Next we deduce the phase space Lagrangian for the guiding center CCS. More specifically, we would like to find a functional of the form
\begin{align}
L_{\text{CCS}}(\boldsymbol{\eta},\bu,\boldsymbol{z},\dot{\boldsymbol{\eta}},\dot{\bu},\dot{\boldsymbol{z}};\rho_0,f_0,\mathbf{A}_0),
\end{align}
with an affine dependence on $\dot{\boldsymbol{\eta}},\dot{\bU},$ and $\dot{\boldsymbol{z}}$ such that the following statement is true: the variational principle
\begin{align}
\delta\int_{t_1}^{t_2}L_{\text{CCS}}(\boldsymbol{\eta},\bu,\boldsymbol{z},\dot{\boldsymbol{\eta}},\dot{\bu},\dot{\boldsymbol{z}};\rho_0,f_0,\mathbf{A}_0)\,\mathrm{d}t=0,
\end{align}
with $\boldsymbol{\eta},\bu,$ and $\boldsymbol{z}$ subjected to arbitrary variations with fixed endpoints, reproduces the guiding center CCS. The name ``phase space Lagrangian'' is motivated by the fact that the guiding center CCS may be cast as a first-order ODE on the infinite-dimensional space consisting of tuples $(\boldsymbol{\eta},\bu,\boldsymbol{z})$. It has been shown in \cite{Burby_thesis} using a method inspired by \cite{Marsden98} that there is a systematic procedure for deriving the phase space Lagrangian for any system with an Euler-Poincar\'e variational principle as soon as the structure of the initial value problem is understood. We refer to \cite{Burby_thesis} for a detailed description of this procedure, which involves a careful study of the (temporal) boundary terms arising from the variational principle. When this procedure is applied to the guiding center CCS, we find that the appropriate phase space Lagrangian is given by
\begin{align}\label{phase_space_action}
L_{\text{CCS}}=&\int\rho\, \bu\cdot\mathbf{V}\mathrm{d}^3x-\int q_hn_h\,\mathbf{A}\cdot\mathbf{V}\,\de^3x+\int_\mu f\, \left[q_h\mathbf{A}+m_h v_\parallel\boldsymbol{b}\right]\cdot\bw \,\de^4 z\nonumber\\
&-\mathcal{H}_{\text{CCS}}(\bw,\rho,f,\mathbf{A}),
\end{align}
where now $\mathbf{V}$ is given in terms of the bulk fluid configuration map as $\dot{\boldsymbol{\eta}}(\mathbf{x}_0)=\mathbf{V}(\boldsymbol{\eta}(\mathbf{x}_0))$, $\bcalX=({\bw},a_\parallel)$ is defined as it was in equation~\eqref{CCS_EP_VP}, and the quantities $\rho,f,$ and $\mathbf{A}$ are again given in terms of $\boldsymbol{\eta}$ and $\boldsymbol{z}$ according to equation~\eqref{pullbacks}. We stress that $\bu$ plays the role of an independent generalized coordinate in $L_{\text{CCS}}$; we do \emph{not} assume $\bu= \mathbf{V}$ when defining $L_{\text{CCS}}$. The CCS Hamiltonian is given by
\begin{align}
\mathcal{H}_{\text{CCS}}=&\frac{1}{2}\int \rho\,|\bu|^2\,\mathrm{d}^3x+\int \rho\,\mathcal{U}(\rho)\,\de^3x\nonumber\\
&+\frac{1}{2\mu_0}\int |\mathbf{B}|^2\,\mathrm{d}^3x+\int_\mu \left(\frac{m_h}{2}v_\parallel^2+\mu B\right)\,f\,\de^4 z.
\end{align}
It is straightforward to show directly that the variational principle associated with $L_{\text{CCS}}$ reproduces the guiding center CCS.

The penultimate step in the derivation is to exploit the fact that, in the phase space variational principle \eqref{phase_space_action}, $\bu$ plays the role of a generalized coordinate in the usual sense \cite{Landau}. We may therefore apply a point transformation to $\bu$ without changing the form of the variational principle. In particular, if we pass into the new system of generalized coordinates defined by the transformation $(\boldsymbol{\eta},\bu,\boldsymbol{z})\mapsto (\boldsymbol{\eta},\bU,\boldsymbol{z})$, where
\begin{align}
\bU=\bu+\rho^{-1}\boldsymbol{b}\int_\mu m_h v_\parallel f\,dv_\parallel
\end{align}
is the total kinetic momentum divided by the bulk mass density, i.e. $\mathbf{M}/\rho$, then
\begin{align}
\delta\int_{t_1}^{t_2}\bar{L}_{\text{CCS}}(\boldsymbol{\eta},\bU,\boldsymbol{z},\dot{\boldsymbol{\eta}},\dot{\bU},\dot{\boldsymbol{z}};\rho_0,f_0,\mathbf{A}_0)\,\mathrm{d}t=0,
\end{align}
is a valid variational principle for the transformed guiding center CCS model, where $\bar{L}_{\text{CCS}}$ is the transformed CCS phase space Lagrangian and $\bU$ is subjected to arbitrary variations. A trivial direct calculation shows that $\bar{L}_{\text{CCS}}$ is given by
\begin{align}
\bar{L}_{\text{CCS}}=&\int \rho\,\bU\cdot\mathbf{V}\,\mathrm{d}^3x-\int_\mu f m_h v_\parallel \boldsymbol{b}\cdot\mathbf{V}\,\de^4 z-\int q_h n_h \mathbf{A}\cdot\mathbf{V}\,\de^3x\nonumber\\
&+\int_\mu f\left[q_h\mathbf{A}+m_h v_\parallel\boldsymbol{b}\right]\cdot\bw\,\mathrm{d}^4{z}-\overline{\mathcal{H}}_{\text{CCS}}(\bU,\rho,f,\mathbf{A}),
\end{align}
where the transformed Hamiltonian is given by
\begin{align}
\overline{\mathcal{H}}_{\text{CCS}}(\bU,\rho,f,\mathbf{A})=\mathcal{H}_{\text{CCS}}\left(\bU-\rho^{-1}\boldsymbol{b}\int_\mu m_hv_\parallel f\,dv_\parallel, \rho, f, \mathbf{A}\right).
\end{align}

We conclude the derivation as follows. Because we assume that the kinetic population is rarefied, we have
\begin{align}
\mathcal{H}_{\text{CCS}}\left(\bU-\rho^{-1}\boldsymbol{b}\int_\mu m_hv_\parallel f\,dv_\parallel, \rho, f, \mathbf{A}\right)\approx \mathcal{H}_{\text{CCS}}(\bU,\rho,f,\mathbf{A}).
\end{align}
This motivates us to mimic the key step in the Hamiltonian derivation of the Vlasov PCS in \cite{Tronci2010} by defining
\begin{align}
 \mathcal{H}_{\text{PCS}}(\bU,\rho,f,\mathbf{A})\equiv \mathcal{H}_{\text{CCS}}(\bU,\rho,f,\mathbf{A}),
\end{align}
to be the PCS Hamiltonian, and
\begin{align}
L_{\text{PCS}}=&\int \rho\,\bU\cdot\mathbf{V}\,\mathrm{d}^3x-\int_\mu f m_h v_\parallel \boldsymbol{b}\cdot\mathbf{V}\,\de^4 z-\int q_h n_h \mathbf{A}\cdot\mathbf{V}\,\de^3x\nonumber\\
&+\int_\mu f\left[q_h\mathbf{A}+m_h v_\parallel\boldsymbol{b}\right]\cdot\bw\,\mathrm{d}^4{z}-\mathcal{H}_{\text{PCS}}(\bU,\rho,f,\mathbf{A}),
\end{align}
to be the PCS phase space Lagrangian.
Our putative guiding center PCS model is then defined by the variational principle
\begin{align}\label{PSL_VP}
\delta\int_{t_1}^{t_2}L_{\text{PCS}}(\boldsymbol{\eta},\bU,\boldsymbol{z},\dot{\boldsymbol{\eta}},\dot{\bU},\dot{\boldsymbol{z}};\rho_0,f_0,\mathbf{A}_0)\,\mathrm{d}t=0,
\end{align}
where $\boldsymbol{\eta},\bU,$ and $\boldsymbol{z}$ are subjected to arbitrary variations with fixed endpoints. If the Euler-Lagrange equations associated with this variational principle reproduce the conservative guiding center PCS model presented in Section \ref{sec:newmod}, we will have succeeded in proving the validity of this alternative derivation. 

We can see that the variational principle \eqref{PSL_VP} does indeed reproduce the conservative guiding center PCS model \eqref{GCPCS-MHD1}-\eqref{GCPCS-MHD3} as follows. In the action principle \eqref{PSL_VP}, variations of $\boldsymbol{z}$, $\bU$, and $\boldsymbol{\eta}$ lead to the conditions
\begin{gather}
q_h\mathbf{E}^*_{\text{PCS}}+q_h\bw\times\mathbf{B}^*-m_h a_\parallel\boldsymbol{b}=0\,,\label{PCS_EL_1}\\
m_h \boldsymbol{b}\cdot\bw-m_h\boldsymbol{b}\cdot\mathbf{V}-\partial_{v_\parallel}\frac{\delta\mathcal{H}_{\text{PCS}}}{\delta f}=0\,,\\
\rho\mathbf{V}-\frac{\delta\mathcal{H}_{\text{PCS}}}{\delta\bU}=0\,,\\
\rho\bigg(\partial_t\bu_{\text{PCS}}+(\nabla\times\bu_{\text{PCS}})\times\mathbf{V}+\nabla(\bu_{\text{PCS}}\cdot\mathbf{V})\bigg)+\rho\nabla\left(\frac{\delta\mathcal{H}_{\text{PCS}}}{\delta\rho}-\bU\cdot\mathbf{V}\right)\nonumber\\
=\mathbf{J}_{\text{PCS}}\times\mathbf{B}-\mathbf{A}\nabla\cdot\frac{\delta\mathcal{H}_{\text{PCS}}}{\delta\mathbf{A}}\,,\label{PCS_EL_3}
\end{gather}
respectively, where $\mathbf{B}^*$ is given in equation~\eqref{EffFields}, and
\begin{align}
\mathbf{J}_{\text{PCS}}=&\frac{\delta\mathcal{H}_{\text{PCS}}}{\delta\mathbf{A}}-q_h \int_\mu(\bw-\mathbf{V})f\,\mathrm{d}v_\parallel-\nabla\times\int_\mu\frac{m_hv_\parallel}{B}(\bw-\mathbf{V})_\perp f\,\mathrm{d}v_\parallel\,,\\
\bu_{\text{PCS}}=&\bU-\rho^{-1}\boldsymbol{b}\int_\mu m_h v_\parallel f\,\mathrm{d}v_\parallel\,,\\
\mathbf{E}^*_{\text{PCS}}=&-\mathbf{V}\times\mathbf{B}-\frac{m_h}{q_h}\frac{v_\parallel}{B}\nabla\times(\mathbf{V}\times\mathbf{B})_\perp-\frac{1}{q_h}\nabla\left(\frac{\delta\mathcal{H}_{\text{PCS}}}{\delta f}+ m_h v_\parallel \boldsymbol{b}\cdot\mathbf{V}\right),
\end{align}
is convenient shorthand notation. By making use of the functional derivative relations
\begin{gather}
\frac{\delta\mathcal{H}_{\text{PCS}}}{\delta f}=\mu B + \frac{m_h}{2}v_\parallel^2\,,\quad\quad \frac{\delta\mathcal{H}_{\text{PCS}}}{\delta \bU}=\rho\bU\,,\quad\quad \frac{\delta\mathcal{H}_{\text{PCS}}}{\delta \rho}=\frac{1}{2}|\bU|^2+\mathcal{U}+\rho\,\mathcal{U}^\prime\,,\\
\frac{\delta\mathcal{H}_{\text{PCS}}}{\delta \mathbf{A}}=\nabla\times\left(\mu_0^{-1}\mathbf{B}+\boldsymbol{b}\int_\mu \mu f\,\mathrm{d}v_\parallel\right),
\end{gather}
and applying the following non-trivial intermediate consequence of the Euler-Lagrange equations:
\begin{align}
\partial_t\left(\boldsymbol{b}\int_\mu m_h v_\parallel f\,\mathrm{d}v_\parallel\right)=&q_h \left(\int_\mu (\bw-\mathbf{V})f\,\mathrm{d}v_\parallel\right)\times\mathbf{B}+\nabla\boldsymbol{b}\cdot m_h \int_\mu v_\parallel\bw_\perp f\,\mathrm{d}v_\parallel\nonumber\\
&-\frac{\nabla B}{B}\int_\mu \mu B f\,\mathrm{d}v_\parallel-\nabla\cdot\left( m_h \int_\mu v_\parallel\bw\boldsymbol{b} f \mathrm{d}v_\parallel\right),
\end{align}
it is straightforward to show that equations~\eqref{PCS_EL_1}-\eqref{PCS_EL_3} recover \eqref{GCPCS-MHD1}-\eqref{GCPCS-MHD3}.

\subsection{Discussion and conservation laws\label{sec:discussion}}

The conservative PCS scheme \eqref{GCPCS-MHD1}-\eqref{GCPCS-MHD3} differs from other PCS schemes appearing in the literature in many ways. Probably, the most direct comparison is with the analogue scheme proposed in Section III.C of \cite{PaBeFuTaStSu}. Although the exact expression of the pressure tensor is not specified in that work, the corresponding guiding center trajectories  can be immediately compared to our kinetic equation \eqref{GCPCS-MHD3}. We can report two main differences: the first is that in \cite{PaBeFuTaStSu} the authors used the approximation $B^*_\|\simeq B$, while the second more important difference is that the inertial force terms involving the bulk fluid velocity in \eqref{GCPCS-MHD3} are totally absent. This is the conventional approach used in the literature to produce PCS models, which destroys the correct energy balance, as we discussed earlier on.

It would {also} be tempting to compare the PCS system \eqref{GCPCS-MHD1}-\eqref{GCPCS-MHD3} with other work on the same topic. However, {many} variants have been proposed  and a direct comparison is not always clear. For example, in the works \cite{Kim,KimEtAl,TaBrKi} the authors use the definition of pressure involving the relative velocity, that is $\widetilde{\Bbb{P}}_h=m_h\int\left(\bv-\boldsymbol{V}\right)\left(\bv-\boldsymbol{V}\right)f\,\de^3v$. As shown in \cite{Tronci2010,HoTr2011} by using full Vlasov kinetics, this form of the pressure tensor can be obtained in an energy-conserving model by decomposing the total energy of the energetic particle and then discarding the mean flow terms while retaining temperature terms. While this approximation is physically very appealing, the resulting energy-conserving equations are much more complicated, which is why we decided to retain the full stress tensor ${\Bbb{P}}_h=m_h\int\bv\bv f\,\de^3v$ in the equations of motion. However, inertial force terms still appear in the energetic particle motion \cite{Tronci2010,HoTr2011}, and these forces are completely absent in the literature.

{Elsewhere} \cite{Cheng,FuPark,FuParketAl,BrVlZo,BrVlZoKa}, variants of the CGL stress tensor $\Bbb{P}_{\tiny\rm CGL}=m_h
\int_\mu \!\big[v_\|^2\bb\bb+\mu B(\boldsymbol{1}-\bb\bb)
\big]f\,
\de v_\|$ are retained in the fluid momentum equation and the energetic particles are described by gyrokinetics. While this case is not treated in this paper (see \cite{BuTr} for conservative  CCS models using gyrokinetics), we may still comment that inertial forces would still be expected to emerge in the gyrokinetic formalism for fully nonlinear conservative PCS models.

The system \eqref{GCPCS-MHD1}-\eqref{GCPCS-MHD3} has been constructed within the Euler-Poincar\'e variational framework by exploiting the coupling between fluid and phase space paths and their Eulerian variable counterparts. As we anticipated, energy conservation follows by construction for the following energy functional (Hamiltonian):
\[
E=\frac12\int\!\rho|\bU|^2\,\de^3x+\iint_{\mu}\!\left(\frac{m_h}2v_\|^2+\mu B \right)f\,\de v_\|\,\de^3x+\int\!\rho\,\mathcal{U}(\rho)\,\de^3{x}+\frac1{2\mu_0}\int\! |\bB|^2\,\de^3x
\,.
\]
This is the sum of the bulk MHD total energy and the energy of the guiding center ensemble for the energetic particles. Notice that this coincides with the same energy functional that is conserved by the CCS system \eqref{cc-hybrid-momentum-GC}-\eqref{cc-hybrid-end-GC}.

Other than energy conservation, the PCS model \eqref{GCPCS-MHD1}-\eqref{GCPCS-MHD3} also enjoys cross-helicity conservation, as it emerges from direct application of equation \eqref{KM-momap}. To see this, we recall the definition $\bK_{\rm gc}=m_h\int_\mu(\bb\cdot\bv)\hat{f}\,\de^3v$ and write \eqref{KM-momap} more explicitly as
\begin{multline*}
\left(\frac{\partial}{\partial t}+\pounds_{\bU}\right)\left(\rho\bU-\bK_{\rm gc}-q_h n_h \bA\right)
=
\rho\,\nabla\left[\frac{U^2}{2}-\mathcal{U}-\rho\mathcal{U}'\right]
\\
+(\bJ-\bJ_{\rm gc}-\nabla\times\bM_{\rm gc})\times\bB
+\left(\nabla\cdot {\bJ}_{\rm gc}\right)\bA
\,,
\end{multline*}
that is, using \eqref{EP-PCS3} and $q_h\partial_tn_h+\nabla\cdot(q_hn_h\bU+\bJ_{\rm gc})=0$,
\begin{multline}
\left(\frac{\partial}{\partial t}+{\bU}\cdot\nabla+\nabla\bU\cdot\right)\left(\bU-\rho^{-1}\bK_{\rm gc}\right)
=
\,\nabla\!\left[\frac{U^2}{2}-\mathcal{U}-\rho\mathcal{U}'\right]
\\
+\rho^{-1}(\bJ-\bJ_{\rm gc}-\nabla\times\bM_{\rm gc})\times\bB
\,.
\label{GCKM-momap2}
\end{multline}
Then, by dotting equation \eqref{GCKM-momap2} with $\bB$ and integrating, we obtain conservation of the following cross-helicity
\beq\label{gcCH}
\frac{\de}{\de t}\int \bB\cdot\left(\bU-\rho^{-1}\bK_{\rm gc}\right)\de^3x=0
\,,
\eeq
which is the guiding center analogue of the full-orbit Vlasov correspondent \eqref{CH}. In addition, the standard frozen-in condition also ensures conservation of magnetic helicity. We remark that, while the latter is also preserved by the non-conservative models appearing in the literature, cross-helicity conservation is a specific feature of the new PCS model \eqref{GCPCS-MHD1}-\eqref{GCPCS-MHD3} and it is lost in other models. As a final remark, we notice that cross-helicity can be exploited to perform Lyapunov stability studies \cite{HoMaRaWe} along the lines of those already carried out for the case of the Vlasov PCS model \eqref{PCS-MHD1}-\eqref{PCS-MHD3} \cite{MoTaTr2014-1,MoTaTr2014}.

\section{Conclusions}

Within the Euler-Poincar\'e variational framework, this paper has formulated the first conservative fully nonlinear PCS model \eqref{GCPCS-MHD1}-\eqref{GCPCS-MHD3} for energetic particles undergoing guiding center motion. This hybrid model was obtained in two ways: first by exploiting the geometric structure underlying the coupling between the Lagrangian paths for the bulk fluid and the energetic particles, thereby extending the approach to PCS modeling given in  \cite{HoTr2011}; and second by applying an Eulerian momentum shift within the CCS phase space Lagrangian, thereby extending the complementary approach to PCS modeling given in \cite{Tronci2010}. This coupling corresponds to a (Lagrangian) frame change which in turn reflects an intricate variational structure for the Eulerian coordinates. This variational structure is dealt with naturally by Euler-Poincar\'e reduction theory. 

The physical effects of the Lagrangian frame change are found in the appearance of specific inertial force terms in the guiding center trajectories. These inertial force terms are essential to guarantee energy conservation, {and confer} physical consistency on the model.

Other than conserving magnetic helicity and energy, the new model \eqref{GCPCS-MHD1}-\eqref{GCPCS-MHD3} also conserves the modified cross-helicity \eqref{gcCH}, a feature that emerges from the rich geometric structure underlying the model. As anticipated, cross-helicity conservation allows for a Lyapunov stability study by the energy-Casimir method \cite{HoMaRaWe,MoTaTr2014-1,MoTaTr2014}.

The variational structure presented in this paper was shown in \cite{Tronci2010} to be accompanied by a specific type of Poisson brackets, which are not treated in this paper. Indeed, when energetic particles are treated by the guiding center approximation, the essentially non-canonical features of this approximation {endow the Poisson brackets with} a high level of complication that has recently been explored in \cite{BuBrMoQi} within the context of gyrokinetic theory and in \cite{BuSe} in the context of kinetic MHD. {Alternatively,} the PCS phase space Lagrangian introduced in Section \ref{sec:josh_sec} bridges the gap between the Lagrangian and Hamiltonian formalisms. Indeed, by exploiting the proximity of the phase space Lagrangian to the Hamiltonian picture, we have successfully applied the Poisson bracket approach to PCS modeling in \cite{Tronci2010} without the necessity of deriving explicit expressions for the cumbersome guiding center PCS Poisson bracket. 

The extension of the PCS model \eqref{GCPCS-MHD1}-\eqref{GCPCS-MHD3} to the case of gyrokinetic (as opposed to guiding center) theory is certainly possible. For the case of CCS models, we refer the reader to \cite{BuTr}. In the case of the PCS, this extension is left for future research. It is expected that similar inertial force terms will also appear in this context and this could be an opportunity to modify existing codes in such a way that they can reflect the physical effects of energetic particle kinetics.

\subsection*{Acknowledgements} We are grateful to Alain J. Brizard for stimulating conversation throughout this work. ARDC acknowledges financial support from the Engineering \& Physical Sciences Research Council Grant ref.  EP/M506655/1, while CT acknowledges that of the Leverhulme Trust Research Project Grant ref. 2014-112. This research was also supported by the U. S. Department of Energy, Office of Science, Fusion Energy Sciences under  Award No. DE-FG02-86ER53223  and the U.S. Department of Energy Fusion Energy Sciences Postdoctoral Research Program administered by the Oak Ridge Institute for Science and Education (ORISE) for the DOE. ORISE is managed by Oak Ridge Associated Universities (ORAU) under DOE contract number DE-AC05-06OR23100. All opinions expressed in this paper are the author's and do not necessarily reflect the policies and views of DOE, ORAU, or ORISE.

\appendix

\section{Appendix}

\subsection{The first guiding-center moment: proof of equation \eqref{gc-K}\label{app1}}
{
Denote the components of the particle vector field that transports $f$ by $\bcalX=(\bw,a_\parallel)$.
Then computing the time derivative of $\bK$ directly, we have
\begin{align*}
	\frac{\pa\bK_{\rm gc}}{\pa t}
	&=m_h\frac{\pa}{\pa t}\int_\mu f\vp\bb \de\vp
	\\
	&=-m_h\int_\mu\vp\nabla\cdot(\bcalX f)\de\vp
	+m_h\int_\mu f\vp\frac{\pa\bb}{\pa t}\de\vp
	\\
	&=-m_h\nabla\cdot\int f\vp\bw\bb^T \de\vp
	+m_h\int_\mu f\vp\bw\cdot\nabla\bb \de\vp
	+m_h\int_\mu f a_\parallel\bb \de\vp
	+m_h\int_\mu f\vp\frac{\pa\bb}{\pa t}\de\vp
\end{align*}
At this point we recall the Euler-Poincar\'{e} equation derived from the CCS Lagrangranian under variations $\delta\bcalX$ (see Section 3.2 of~\cite{BuTr}),
\begin{align*}
	a_\parallel\bb&=a_h(\bw\times\bB^*+\bE^*)
	\\
	&=a_h(\bw\times\bB^*
	-\bU\times\bB-a_h^{-1}v_\parallel \partial_t\bb-q_h^{-1}\mu\nabla B)\,,
\end{align*}
where we have use the expression for $\bE^*$ in~\eqref{EffFields}. Replacing this in the equation for $\pa_t\bK_{\rm gc}$ above, and using $\bw\cdot\nabla\bb-\bb\cdot\nabla\bw=\nabla\times\bb\times\bw$, we have
\begin{align*}
	\frac{\pa\bK_{\rm gc}}{\pa t}
	&=-m_h\nabla\cdot\int_\mu f\vp\bw\bb^T \de\vp
	+m_h\int_\mu f\vp\nabla\bb\cdot\bw \de\vp
	\\&\hspace{5cm}
	+\int_\mu f\Big(q_h(\bw-\bU)\times\bB-\mu\nabla B\Big)\de\vp\,,
\end{align*}
We then recognize that
\[
	\int_\mu f\Big(m_h\vp\nabla\bb\cdot\bw -\mu\nabla B\Big)\de\vp
	=\nabla\bB\cdot\bM_{\rm gc}
	=\nabla\times\bM_{\rm gc}\times\bB
	+\nabla\cdot\Big((\bB\cdot\bM_{\rm gc})\boldsymbol{1}
	-\bB\bM^T_{\rm gc}\Big)\,,
\]
for $\bM_{\rm gc}$ defined in~\eqref{magn-corr}. Then, with $n_h=\int_\mu f\de\vp$ and $\bJ_{\rm gc}=\int_\mu q_h f\bw \de\vp$, we arrive at 
\begin{align*}
	\frac{
	\pa\bK_{\rm gc}}{\pa t}&=-\nabla\cdot\int_\mu fm_h\vp\bw\bb^T \de\vp
	+\nabla\cdot\Big((\bB\cdot\bM_{\rm gc})\boldsymbol{1}
	-\bB\bM^T_{\rm gc}\Big)
	+\Big(\bJ_{\rm gc}+\nabla\times\bM_{\rm gc}-n_hq_h\bU\Big)\times\bB
	\\
	&=-\nabla\cdot\int_\mu f\Big(m_h\vp^2\bb\bb^T
	+m_h\vp\bw_\perp\bb^T\Big)\de\vp
	-\nabla\cdot\int_\mu f\Big(m_h\vp\bb\bw^T_\perp
	+\mu B(\boldsymbol{1}-\bb\bb^T)\Big)\de\vp
	\\&\hspace{4cm}
	+\Big(\bJ_{\rm gc}+\nabla\times\bM_{\rm gc}-n_hq_h\bU\Big)\times\bB
	\\
	&=-\nabla\cdot\Bbb{P}_{\rm gc}
	+\Big(\bJ_{\rm gc}+\nabla\times\bM_{\rm gc}-n_hq_h\bU\Big)\times\bB\,.
\end{align*}
}

\subsection{Proof of equation \eqref{lemma} and the Lagrange-to-Euler map\label{app2}}
{
We express the tangent lift of the decomposed coordinates $\z=(\bx,\bv)=(\bx,v_\parallel,\bv_\perp)$ by
\[
	\mathcal{T}\!\et(\bx,v_\parallel,{\bf v}_\perp)
	=\Big(\et({\bf x}),{\bf v\cdot\nabla\et(x)\cdot \bb(\et(x})),
	\Pi_\perp({\bf v\cdot\nabla\et(x}))\Big)\,,
\]
where $\Pi_\perp=\boldsymbol{1}-\bb(\et(\bv))\bb(\et(\bv))^T$ is the perpendicular projection operator.
To see how the projected component $\int\hat f({\bf x},\vp,{\bf v}_\perp)d{\bf v}_\perp=f({\bf x},\vp)$ is advected, we use the Lagrange-to-Euler map when taking its time derivative. Let $\Psi:=\mathcal{T}\!\et\circ\ps$. Then,
\begin{align*}
	\frac{\pa}{\pa t}\int{\hat f}\de^2{ v}_\perp
	&=\frac{\pa}{\pa t}\int{\hat f}_0
	\delta\Big({\bf z}-\Psi(\bz_0)\Big)
	\de^6z_0\de^2v_\perp
	\\
	&=\frac{\pa}{\pa t}\int{\hat f}_0
	\delta\Big({\bf x}-\Psi_{\bf x}(\bz_0)\Big)
	\delta\Big(v_\parallel-\Psi_{v_\parallel}(\bz_0)\Big)
	\delta\Big(\bv_\perp-\Psi_{\bv_\perp}(\bz_0)\Big)
	\de^6z_0\de^2v_\perp
	\\
	&=-\nabla\cdot\int{\hat f}_0
	\delta\Big({\bf z}-\Psi(\bz_0)\Big)
	\frac{\pa\Psi_{\bf x}}{\pa t}({\bf z}_0)
	\de^6z_0\de^2v_\perp
	\\&\hspace{1cm}
	-\frac{\pa}{\pa v_\parallel}\int{\hat f}_0
	\delta\Big({\bf z}-\Psi(\bz_0)\Big)
	\frac{\pa\Psi_{v_\|}}{\pa t}({\bf z}_0)
	\de^6z_0\de^2v_\perp
	\\&\hspace{1cm}
	-\frac{\pa}{\pa\bv_\perp}\int{\hat f}_0
	\delta\Big({\bf z}-\Psi(\bz_0)\Big)
	\frac{\pa\Psi_{\bv_\perp}}{\pa t}({\bf z}_0)
	\de^6z_0\de^2v_\perp\,,
\end{align*}
where $\de^6z_0=\de^3x_0\de {\vp}_0{\de^2v_\perp}_0$ and the subscripts ${\bf x},\vp$, and ${\bf v}_\perp$ denote the respective components of the maps. The final term vanishes after integration over $\bv_\perp$, leaving us with
\[
	\frac{\pa}{\pa t}\int{\hat f}\de^2{v}_\perp
	=-\nabla\cdot\int{\hat f}
	\frac{\pa\Psi_{\bf x}}{\pa t}\Big|_{\Psi^{-1}}\de^2v_\perp
	-\frac{\pa}{\pa v_\parallel}\int{\hat f}
	\frac{\pa\Psi_{v_\|}}{\pa t}\Big|_{\Psi^{-1}}\de^2v_\perp\,,
\]
where the notation $|_{\Psi^{-1}}$ means evaluated at (i.e., composed with) $\Psi^{-1}$. By inspection, then, the $\bf x$-component of the advecting vector field is
\begin{align*}
	\hat{\bcalX}_\bx:&=\frac{\pa\Psi_{\bf x}}{\pa t}\Big|_{\Psi^{-1}}
	\\
	&=\frac{\pa\mathcal{T}\!\et_{\bf x}}{\pa t}
	\Big|_{\mathcal{T}\!\et^{-1}}
	+\Big(\frac{\pa \mathcal{T}\!\et}{\pa\ps}\cdot\dot\psi\circ\psi^{-1}\Big)_{\bf x}
	\Big|_{\mathcal{T}\!\et^{-1}}
	\\
	&=\Big[\bcalX_\bU+\bcalX\Big]_{\bf x}
	\\
	&=\bU+\bw
	\\
	&=\left[\bb\cdot(\bv+ \bU)\right]\frac{\bB^*}{B^*_\|}-\frac{\bb}{B^*_\|}
	\times\left[\bE^*-a_h^{-1}
	\left(\bb\cdot\bv\right)\nabla(\bb\cdot\bU)\right]\,,
\end{align*}
where the last equality follows from~\eqref{u-eq}.
Then finally, we use
\[
	(\mathcal{T}\!\et\circ\psi)_{v_\parallel}(\bz_0)
	=(\mathcal{T}\!\et\circ\psi)_{\bf v}(\bz_0)
	\cdot\bb((\mathcal{T}\!\et\circ\psi)_{\bf x}(\bz_0))\,,
\]
whose time-derivative composed from the right with $(\mathcal{T}\!\et\circ\psi)^{-1}$ gives us the following.
\begin{align*}
	\hat{\bcalX}_\vp
	&:=\frac{\pa\Psi_\vp}{\pa t}\Big|_{\Psi^{-1}}
	\\
	&=\frac{\pa\Psi_\bv}{\pa t}
	\Big|_{\Psi^{-1}}\cdot\bb(\Psi_{\bf x})\Big|_{\Psi^{-1}}
	+\Psi_\bv\Big|_{\Psi^{-1}}
	\cdot\frac{\pa\bb(\Psi_{\bf x})}{\pa t}\Big|_{\Psi^{-1}}
	+\frac{\pa\Psi_{\bf x}^i}{\pa t}\Big|_{\Psi^{-1}}
	\frac{\pa\bb^j}{\pa\Psi^i_{\bf x}}\Big|_{\Psi^{-1}}
	\Psi^j_\bv\Big|_{\Psi^{-1}}
	\\
	&=(\bcalX_\bU+\bcalX)_\bv\cdot\bb
	+\bv\cdot\frac{\pa\bb}{\pa t}
	+(\bcalX_\bU+\bcalX)_{\bf x}\cdot\nabla\bb\cdot\bv
	\\
	&=\ba\cdot\bb
	+\bv\cdot\nabla\bU\cdot\bb
	+\bv\cdot\frac{\pa\bb}{\pa t}
	+(\bU+\bw)\cdot\nabla\bb\cdot\bv
	\\
	&=\frac{\bB^*}{B^*_\|}\cdot\left(a_h\bE^*-v_\|\nabla U_\|\right)\,,
\end{align*}
the last equality following from~\eqref{a-eq}. Since neither the $\bf x$- nor the $\vp$-components of the vector field depend on $\bf v_\perp$, we arrive at
\[
	\frac{\pa}{\pa t}\int_\mu{\hat f}\de^2v_\perp
	+\nabla\cdot\left(\hat{\bcalX}_\bx\int_\mu{\hat f}\de^2v_\perp\right)
	+\frac{\pa}{\pa\vp}\left({\hat \bcalX}_\vp\int_\mu{\hat f}\de^2v_\perp\right)
	=0\,,
\]
and~\eqref{lemma} holds.
}

\subsection{Proof of equation \eqref{Jack}\label{app3}}
{
Using the variational derivatives of Section~\ref{sec:newmod},
\[
	\frac{\delta l}{\delta\bcalX}
	=\Big(\hat{f}q_h\bA^*\,,\,0\Big)\,,
	\qquad
	\frac{\delta l}{\delta \hat{f}}
	=q_h\bA^*\cdot\bw-\frac{m_h}2(\bb\cdot\bv)^2-q_h^{-1}\mu B\,,
\]
then we have
\begin{align*}
	\int_\mu\left(\pounds_\bcalX\frac{\delta l}{\delta\bcalX}
	-f\nabla_\bz\frac{\delta l}{\delta \hat{f}}\right)_{\!\!\bx}\de^3v
	&=\int_\mu\left(\nabla\cdot\Big(q_h\hat{f}\bw\bA^*\Big)
	+q_h\hat{f}\nabla\bu\cdot\bA^*\right)\de^3v
	\\&\hspace{1.5cm}
	-\int_\mu\hat{f}\Big(\nabla(q_h\bA^*\cdot\bu)
	-m_h(\bb\cdot\bv)\nabla(\bb\cdot\bv)
	-\mu\nabla B\Big)\de^3v
	\\
	&=\nabla\cdot\int_\mu\Big(q_h\hat{f}\bw\bA^*\Big)\de^3v
	-\int_\mu q_h\hat{f}\nabla\bA^*\cdot\bu\de^3v
	\\&\hspace{1.3cm}
	+\int_\mu\hat{f}\Big(
	m_h(\bb\cdot\bv)\nabla(\bb\cdot\bv)
	+\mu\nabla B\Big)\de^3v
	\\
	&=\nabla\cdot\int_\mu\Big(q_h\hat{f}\bw\bA^*\Big)\de^3v
	-\int_\mu q_h\hat{f}\nabla\bA\cdot\bu\de^3v
	\\&\hspace{1.3cm}
	+\int_\mu\hat{f}\Big(\mu\nabla B
	-m_h(\bb\cdot\bv)\nabla\bb\cdot\bu)\Big)\de^3v
	\\
	&=\nabla\cdot\int_\mu\Big(q_h\hat{f}\bw\bA^*\Big)\de^3v
	-\int_\mu q_h\hat{f}\nabla\bA\cdot\bu\de^3v
	-\nabla\bB\cdot\bM_{\rm gc}\,.
\end{align*}
Meanwhile, 
\begin{align*}
	-\nabla\cdot\!\int\!\bv\left(\!\pounds_{ \bcalX}	\frac{\delta
	\ell}{\delta  \bcalX}-\hat{f}\,
	\nabla_\bz\frac{\delta\ell}{\delta \hat{f}}\right)_{\!\!\bv}\de^3v
	&=-\nabla\cdot\int_\mu\hat{f}\bv\left(
	q_h\frac{\pa\bu}{\pa\bv}\cdot{\bf A}^*
	-q_h\frac{\pa}{\pa\bf v}\Big(
	\bw\cdot{\bf A^*}\Big)
	+m_h\bb\bb\cdot\bv\right)\de^3v
	\\&=0\,,
\end{align*}
since,
\begin{align*}
	m_h(\bv\cdot\bb)\bv\bb^T
	+q_h{\bf v}\frac{\pa\bu}{\pa{\bf v}}\cdot{\bf A}^*
	-q_h{\bf v}\frac{\pa}{\pa\bf v}(\bw\cdot{\bf A^*})=0\,.
\end{align*}

}

\end{document}